\documentclass[twocolumn,amssymb,nobibnotes,showpacs,superscriptaddress,aps,prl,longbibliography]{revtex4-2}

\usepackage{amsmath,graphicx}

\usepackage[colorlinks,linkcolor=blue,urlcolor=blue,anchorcolor=blue,citecolor=blue]{hyperref}

\begin{document}


\title{Strong antibunching photons/photon pairs emission in two atoms cavity QED system with the Van der Waals interaction}


\author{Z. Y. Yu}
\affiliation{MOE Key Laboratory of Advanced Micro-Structured Materials,
	School of Physics Science and Engineering, Tongji University, Shanghai, China 200092}
\author{J. P. Xu}
\affiliation{MOE Key Laboratory of Advanced Micro-Structured Materials,
	School of Physics Science and Engineering, Tongji University, Shanghai, China 200092}
\author{C. J. Zhu}
\email[Corresponding author:]{cjzhu@suda.edu.cn}
\affiliation{School of Physical Science and Technology, Soochow University, Suzhou 215006, China}
%
\author{Z. Y. Bai}
\email[Corresponding author:]{zhybai@nju.edu.cn}
\affiliation{National Laboratory of Solid State Microstructures and School of Physics, 
    Collaborative Innovation Center of Advanced Microstructures, 
    Nanjing University, Nanjing 210093, China}
\author{Y. P. Yang}
\email[Corresponding author:]{yang\_yaping@tongji.edu.cn}
\affiliation{MOE Key Laboratory of Advanced Micro-Structured Materials,
	School of Physics Science and Engineering, Tongji University, Shanghai, China 200092}
	



\date{\today}

\begin{abstract}
	We investigate the generation of antibunching photons and photon pairs in a two-atom cavity QED system leveraging interatomic van der Waals (vdW) interaction. We show that the vdW interaction shifts the two-atom excited state, enabling the suppression of two-photon excitation via destructive interference in a diamond configuration. This leads to antibunching photon emission with extremely high purity. Furthermore, by tuning the vdW interaction strength, the conditions for conventional and unconventional photon blockades can overlap, significantly enhancing single-photon emission. Conversely, at the antiblockade excitation frequency, quantum Monte Carlo simulations demonstrate the feasibility of generating antibunching photon pairs with a high purity and reasonable leaking rate by selecting appropriate driving field Rabi frequency and vdW interaction strength. These findings on enhanced single/two-photon emission could lead to more efficient and brighter quantum light sources.
\end{abstract}

\maketitle

\section{Introduction}
%
Photon blockade (PB) is a quantum mechanical effect which not only leads to a nonclassical state of light with antibunching emission of photons~\cite{zou1990photon,lemonde2014antibunching,ren2021antibunched}, but also serves as a tool to unveil various quantum features at the single-photon level~\cite{chen2022photon,lu2021plasmonic}. Conventional photon blockade (CPB) originates from the anharmonic splitting of eigenstates in cavity quantum electrodynamics (QED) with strong couplings between atoms and cavity~\cite{scully1999quantum,birnbaum2005photon}. It plays a crucial rule in quantum information~\cite{hacker2016photon,tiarks2019photon,li2020photon}, generating non-classical lights~\cite{dayan2008photon,huang2018nonreciprocal,muller2015coherent,faraon2010generation,faraon2008coherent} and building single-photon sources~\cite{flayac2015all,tang2021towards,majumdar2013single}.
Contrary to the CPB, unconventional photon blockade (UCPB) has been proposed in the weak coupling regime~\cite{liew2010single,flayac2017unconventional,zubizarreta2020conventional}. It utilizes quantum destructive interference (DI) of transition pathways to eliminate the two-photon transitions~\cite{bamba2011origin,hou2019interfering,zhou2025universal}. As a result, extremely strong photon blockade behavior can be observed under the condition of UCPB. However, 
although the purity of antibunching photons generated by UCPB is very high, the emission rate of photons is much lower than that generated by CPB. 

Recently, both CPB and UCPB have been extensively studied and experimentally demonstrated in various QED systems, including cavity QED systems~\cite{snijders2018observation,ridolfo2012photon,peyronel2012quantum,radulaski2017photon,
trivedi2019photon,chakram2022multimode,li2022single,jabri2022enhanced,xia2021giant,wang2021giant,shen2020nonreciprocal}, circuit-QED systems~\cite{hoffman2011dispersive,lang2011observation,liu2014blockade,vaneph2018observation}, optomechanical systems~\cite{rabl2011photon,li2019nonreciprocal,liao2013photon,liao2013correlated,komar2013single,sun2023photon,gao2023phase,liu2023nonreciprocal} and so on~\cite{li2024enhancement,zhang2023nonreciprocal,lu2025chiral,xie2022nonreciprocal,fan2024nonreciprocal}. Similar method based on energy level anharmonicity (ELA) has been used to realize two photon blockade, facilitating the resonant absorption of two photons and subsequently reducing the transition probability of the third photon~\cite{miranowicz2013two,hovsepyan2014multiphoton,deng2015enhancement,zhu2017collective,bin2018two,kowalewska2019two,hamsen2017two,zhou2021n}. This quantum scissor can thereby provide the emission of antibunching photon pairs~\cite{singh2019photon,ren2021antibunched}. Furthermore, multiphoton blockade can also be studied via multiphoton transitions~\cite{zhou2021n}. While the third-order correlation function can readily be suppressed (indicating blockade of the third photon), achieving a sufficiently strong two-photon bunching, where the second-order correlation function is substantially greater than unity, remains elusive. This limitation hinders the efficient generation of strongly antibunching photon pairs, which are a key signature of two-photon blockade.

Since the key to achieving PB is to control energy levels and transition probabilities, the interaction between atoms provides a new degree of freedom~\cite{qu2020improving,devi2020nonequilibrium,williamson2020superatom,cidrim2020photon,zhu2021hybrid,zheng2016photon}. In atomic systems, the interatomic dipole-dipole interaction (DDI) and the van der Waals (vdW) interaction are good candidates for manipulating atomic states. The former leads to energy shifts of single atom excitation states, which can be used to realized strong single PB via the destructive interference~\cite{zhu2021hybrid}. However, the latter shifts the energy of the two atoms excitation state and leads to the well-known Rydberg blockade phenomenon, where only one atom can be excited to the Rydberg state~\cite{lukin2001dipole,urban2009observation,gaetan2009observation}. 
In contrast to the Rydberg blockade, a well-detuned laser can compensate the vdW induced energy shift and two atoms can be simultaneously excited to the Rydberg states, which is known as the antiblockade phenomenon~\cite{ates2007antiblockade,amthor2010evidence}. 

In this paper, we investigate how to realize antibunching photons or photon pairs emission via the van der Waals interaction in two atoms cavity QED system. In the presence of the interatomic vdW interaction, the excited state for two atoms is shifted. By choosing a specific driving field frequency, the excitation of two photon state will be forbidden due to the destructive interference effect with a diamond configuration. As a result, strong emission of antibunching photons can be achieved based on the UCPB phenomenon. Adjusting the vdW interaction strength, conditions for the generation of CPB and UCPB overlap and the emission rate of single photon can be enhanced significantly. We also show, at the excitation frequency of antiblockade, bunching photons leaking from the cavity with steady-state second order correlation function $g^{(2)}(0)\gg1$. By choosing suitable driving field Rabi frequency and the vdW interaction strength, we carry out the quantum Monte Carlo (MC) simulation and demonstrate the possibility for the emission of bunching photon pairs with reasonable leaking rate.

\section{Model}
%
\begin{figure}[htb]
	\centering
	\includegraphics[width=\linewidth]{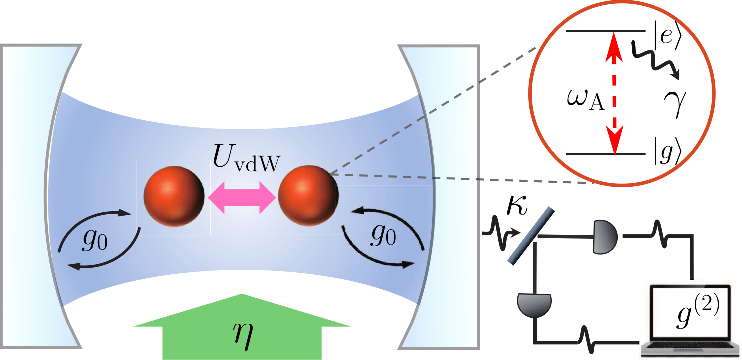}\\
	\caption{Schematic diagram of two identical two-level atoms cavity QED system. Each atom is simplified as a two level scheme with a ground state $|g\rangle$ and an excited state $|e\rangle$. The coupling between each atom and the cavity is labeled by $g_0$, and the van de Waals interaction between two atoms is denoted by $U_{\rm vdW}$. A coherent field with Rabi frequency $\eta$ drives two atoms in phase. The field leaks from the right side of the cavity with the rate $\kappa$, and the excited state of each atom decays with the rate $\gamma$.}~\label{fig:fig1}
\end{figure}
As shown in Fig.~\ref{fig:fig1}, we consider that two identical two-level atoms with the transition frequency $\omega_{\rm A}$ are trapped in a single mode cavity with the resonance frequency $\omega_{\rm cav}$. To realize strong single photon and two photon emissions, two atoms should be prepared close enough to realize the van de Waals interaction, denoted by $U_{\rm vdW}$. Then, we coherently drive these two atoms in phase by an optical field with angular frequency $\omega_{\rm d}$. In the frame rotating with the frequency of the driving field, the Hamiltonian of atoms and cavity is $H_0=\Delta_{\rm cav}a^\dag a+\Delta_{\rm A}(\sigma_+^{(1)}\sigma_-^{(1)}+\sigma_+^{(2)}\sigma_-^{(2)})$ where $a^\dag$ ($a$) is the photon creation (annihilation) operator and $\sigma^{(j)}_\pm$ is the atomic rasing or lowering operator for the $j-$th atom ($j=1-2$). The atom and cavity detunings are defined as $\Delta_{\rm A}=\omega_{\rm A}-\omega_{\rm d}$ and $\Delta_{\rm cav}=\omega_{\rm cav}-\omega_d$, respectively. In our system, the Hamiltonian for the interaction between atoms and cavity is described by $H_I=\sum_jg_j(a^\dag \sigma^{(j)}_-+a\sigma^{(j)}_+)+U_{\rm vdW}\sigma_+^{(1)}\sigma_-^{(1)}\sigma_+^{(2)}\sigma_-^{(2)}$, where the first term represents the atom-cavity interaction with $g_j$ being the coupling strength for the $j-$th atom. The second term indicates the van der Waals interaction between two atoms. Such kind of interaction can be experimentally realized by combining outstanding achievements in the field of cavity QED~\cite{yan2023superradiant,ho2025optomechanical} and Rydberg atom experiments~\cite{browaeys2020many,bluvstein2024logical}. For example, one can use optical photons and Rydberg atoms prepared via the two photon transition or microwave photons and Rydberg atoms initially prepared in a high energy level~\cite{morgan2020coupling}. The Hamiltonian for the driving field is $H_d=\eta\sum_j(\sigma_+^{(j)}+\sigma_-^{(j)})$ with the Rabi frequency $\eta$.

In general, the dynamical behavior of such a system can be described by using the master equation, i.e.,
\begin{eqnarray}\label{eq:ME}
&& \frac{d\rho}{dt}=-i[H_{\rm sys},\rho]+{\cal L}_{\rm atom}(\rho)+{\cal L}_{\rm cav}(\rho),
\end{eqnarray}
where the Hamiltonian of the system is $H_{\rm sys}=H_0+H_I+H_d$. The last two terms in Eq.~(\ref{eq:ME}) represent the decays of atoms and the cavity field, which is defined as ${\cal L}_{\rm atom}(\rho)=\gamma\sum_j(2\sigma_{-}^{(j)}\rho\sigma_{+}^{(j)}-\sigma_{+}^{(j)}\sigma_{-}^{(j)}\rho-\rho\sigma_{+}^{(j)}\sigma_{-}^{(j)})$ and ${\cal L}_{\rm cav}(\rho)=\kappa\sum_j(2a\rho a^\dag-a^\dag a\rho-\rho a^\dag a)$, respectively. The field decay rate is described by $2\kappa$, while the excited state decays with the rate $2\gamma$. 

\begin{figure}[htb]
 \centering
 \includegraphics[width=\linewidth]{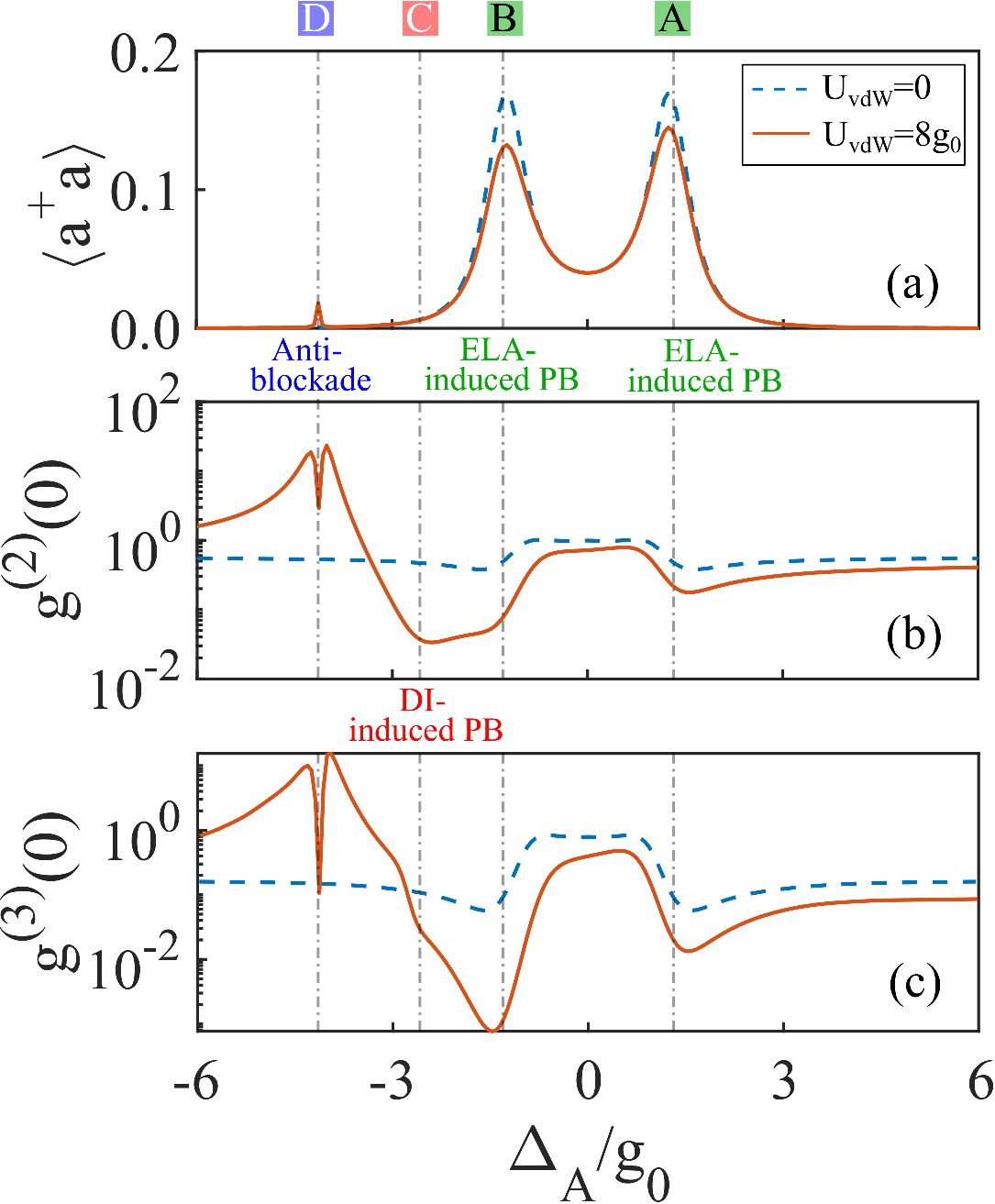}\\
 \caption{(a) Cavity excitation spectrum $\langle a^\dag a\rangle$, (b) the steady state second order correlation function $g^{(2)}(0)$, and (c) the third order correlation function $g^{(3)}(0)$ versus the normalized atomic detuning $\Delta_A/g_0$. The blue dashed curves and the red solid curves in panels (a)-(c) represent the van der Waals interaction strength $U_{\rm vdW}=0$ and $U_{\rm vdW}=8g_0$, respectively. Here, labels A and B denote the frequency for the energy level anharmonicity induced single photon blockade, while labels C denotes the frequency for the destructive interference induced single photon blockade. In addition, label D indicates the frequency for the anti-blockade behavior, which leads to the two photon emission with $g^{(2)}(0)>1$ and $g^{(3)}(0)<1$.~\label{fig:fig2}}
\end{figure}
Without losing the generality, we assume $\Delta_{\rm cav}=\Delta_A$, i.e., $\omega_{\rm cav} =\omega_{\rm_A} $, and $g_1=g_2=g_0$ in the following calculation. Then, we perform a steady-state numerical simulation based on the master equation [i.e., Eq.~(1)] with the inverse power method. In Fig.~\ref{fig:fig2}, the cavity excitation spectrum $\langle a^\dag a\rangle$ [panel (a)], the steady-state second order correlation function $g^{(2)}(0)\equiv\langle a^\dag a^\dag aa\rangle/\langle a^\dag a\rangle^2$ [panel (b)] and the steady-state third order correlation function $g^{(3)}(0)\equiv\langle a^\dag a^\dag a^\dag aaa\rangle/\langle a^\dag a\rangle^3$ [panel (c)] are depicted as a function of the normalized atomic detuning $\Delta_A/g_0$. Here, the van der Waals interaction strength is chosen as $U_{\rm vdW}=0$ (blue dashed curves) and $U_{\rm vdW}=8g_0$ (red solid curves), respectively. Other system parameters are chosen as $g_0=5$ MHz, $\eta=1$ MHz, $\kappa=3$ MHz and $\gamma=5$ kHz, respectively~\cite{browaeys2016interacting,arias2019realization}. To ensure numerical convergence, we truncate the Hilbert space of the cavity mode to a size of $30$ in our numerical simulation. 

In the absence of the van der Waals interaction, i.e, $U_{\rm vdW}=0$, there exist two peaks in the cavity excitation spectrum at the frequencies of $\Delta_A=\pm\sqrt{2}g_0$ (see blue dashed curve), where single photon blockade behavior can be observed with $g^{(3)}(0)<g^{(2)}(0)<1$ (labeled by A and B). The physical mechanism is well known as the energy level anharmonicity induced by the atom cavity coupling, demonstrated in Fig.~\ref{fig:fig3}(a). When the driving field frequency is tuned to be resonant with the $\Psi^{(0)}\leftrightarrow\Psi_\pm^{(1)}$ transition frequency, the $\Psi_\pm^{(1)}\leftrightarrow\Psi_\pm^{(2)}$ transition involving the absorption of the second cavity photon is prohibited due to the energy difference. In the presence of the van der Waals interaction, e.g., $U_{\rm vdW}=8g_0$, one can observe three peaks in the cavity excitation spectrum [see red solid curve in panel (a)] at the frequencies of $\Delta_A=\pm\sqrt{2}g_0$ and $\Delta_A=-U_{\rm vdW}/2$, respectively. It is clear to see that the ELA-induced photon blockade behavior can be improved at $\Delta_A=\pm\sqrt{2}g_0$ [see red curve in Fig.~\ref{fig:fig2}(b)]. The corresponding physical mechanism can also be described by exploring the dressed states shown in Fig.~\ref{fig:fig3}(a). The eigenvalues and eigenstates of the dressed states can be obtained by diagonize the Hamiltonian $H=\omega_{\rm A}(a^{\dag}a+\sigma_+^{(1)}\sigma_-^{(1)}+\sigma_+^{(2)}\sigma_-^{(2)})+\sum_jg_j(a^\dag \sigma^{(j)}_-+a\sigma^{(j)}_+)+U_{\rm vdW}\sigma_+^{(1)}\sigma_-^{(1)}\sigma_+^{(2)}\sigma_-^{(2)}$ in one-photon and two-photon spaces, respectively. As shown in Fig.~\ref{fig:fig3}(a), increasing the van der Waals interaction, only dressed states in two photon space will be affected. As a result, this atomic interaction induced energy shift in two photon space enhances the energy anharmonicity, leading to the improvement of the single photon blockade behavior, specially at the frequency of $\Delta_A=-\sqrt{2}g_0$ (see results labeled by B). In addition, at the frequency of antiblockade regime, i.e., $\Delta_A=-U_{\rm vdW}/2$, two atoms can be excited simultaneously and two photon blockade behavior can be observed with $g^{(2)}(0)>1$ and $g^{(3)}(0)<1$ (denoted by D). More features of two photon emissions will be discussed in section 4.


\section{Interference induced strong single photon emission}
\begin{figure}[htb]
	\includegraphics[width=\linewidth]{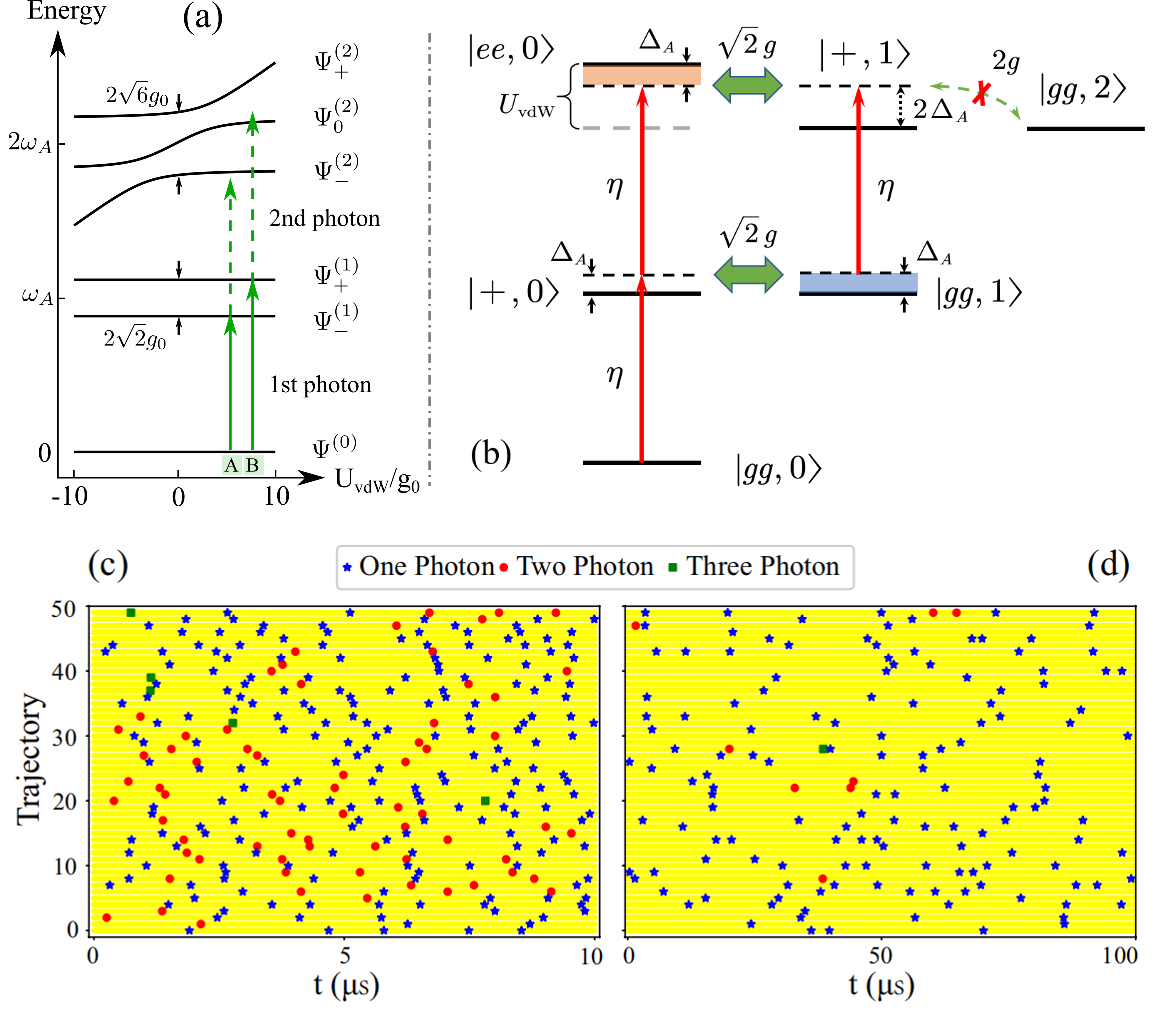}\\
	\caption{(a) Dressed state picture of two atoms cavity QED system with the van der Waals interaction. The green arrows indicate the excitation pathways by absorbing a single photon, while the dashed ones denotes the blockade of the excitation by absorbing the second photon. This EIA induced PB phenomenon leads to $g^{(2)}(0)<1$ labeled by A and B in Fig.~\ref{fig:fig2}. (b) Collective state picture and the corresponding transition pathways of the system. When the detunings of states $|ee,0\rangle$ and $|gg,1\rangle$ have the same magnitude but opposite sign, the destructive interference with a diamond configuration is formed, which prevent the state $|gg,2\rangle$ from being excited. Panels (c) and (d) show the MC simulations of photon emissions with the frequency labeled by A and C, respectively. They record photon clicks when the system undergoes a quantum jump. }~\label{fig:fig3}
\end{figure}
In Fig.~\ref{fig:fig2}, there exists a ``magic" detuning labeled by C, representing the destructive interference induced PB, where the second order correlation function is much smaller than that at the frequencies for ELA-induced PB, i.e., $\Delta_A=\pm\sqrt{2}g_0$. To understand this interesting phenomenon, we use the collective operators to rewrite the Hamiltonian of the system, which yields
\begin{eqnarray}
	H'&=&\Delta_{\rm cav}a^\dag a+\Delta_A J_z+\sqrt{2}g_0(aJ_++a^\dag J_-)\nonumber\\
	& &+\sqrt{2}\eta(J_{+}+J_{-})+\frac{U_{\rm vdW}}{2}(1+J_z)J_z
\end{eqnarray}
where atomic spin collective operators are defined as $J_z=\sigma_z^{(1)}+\sigma_z^{(2)}$, $J_\pm=(\sigma^{(1)}_\pm+\sigma^{(2)}_\pm)/\sqrt{2}$. As discussed in previous section, the last term (Van der Waals interaction) in the 
Hamiltonian $H'$ only affect the state $|ee,n\rangle$, leading to an energy shift of $U_{\rm vdW}$. The collective states and their corresponding transition pathways in one-photon and two-photon spaces are demonstrated in Fig.~\ref{fig:fig3}(b).

%

Clearly, there exist two transition pathways for the excitation of the state $|gg,2\rangle$, i.e., $|gg,0\rangle\overset{\eta}{\rightarrow}|+,0\rangle\overset{\eta}{\rightarrow}|ee,0\rangle\overset{\sqrt{2}g_0}{\rightarrow}|+,1\rangle\overset{2g_0}{\rightarrow}|gg,2\rangle$ and $|gg,0\rangle\overset{\eta}{\rightarrow}|+,0\rangle\overset{\sqrt{2}g_0}{\rightarrow}|gg,1\rangle\overset{\eta}{\rightarrow}|+,1\rangle\overset{2g_0}{\rightarrow}|gg,2\rangle$. Therefore, the destructive interference with a diamond configuration will be formed if the detunings of states $|gg,1\rangle$ and $|ee,0\rangle$ have the same magnitude, but opposite signs, i.e., $U_{\rm vdW}+2\Delta_A=-\Delta_A$. Thus, the ``magic" detuning can be derived as $\Delta_{\rm magic}\equiv\Delta_A=-U_{\rm vdW}/3$~\cite{hou2019interfering}. Under this condition, the excitation of the state $|gg,2\rangle$ can be prohibited due to the destructive interference effect and strong antibunching photons can be observed, i.e., the DI induced PB phenomenon.

Next, we apply a quantum Monte Carlo approach to demonstrate this single PB behavior, where one follows individual trajectories of the system and records photon clicks whenever the system undergoes a quantum jump. Figures \ref{fig:fig3}(c) and (d) present a series of detection events with the detunings $\Delta_A=\sqrt{2}g_0$ (EIA labeled by A) and $\Delta_A=-U_{\rm vdW}/3$ (DI label by C), respectively. The horizontal axis represents time, and each point denotes a single-photon, two-photon or multiphoton emission event. As show in Fig.~\ref{fig:fig3}(c), except for  about $10\%$ unexpected photon emission processes such as two photon and multiphoton processes, most photons leaking from the cavity are antibunching due to ELA induced PB at the detuning $\Delta_A=\sqrt{2}g_0$. However, in the regime of DI induced PB, unexpected photon emissions (such as two- and multi-photon processes) are significantly suppressed due to the destructive interference effect, yielding an approximate $100\%$ antibunching photon emission as shown in Fig. 3(d). Comparing these two figures, we also notice that total photon emission events in panel (d) is less than that in panel (c) since the destructive interference usually results in very small mean photon number, which prevent it from practical applications.

\begin{figure}[htb]
	\includegraphics[width=\linewidth]{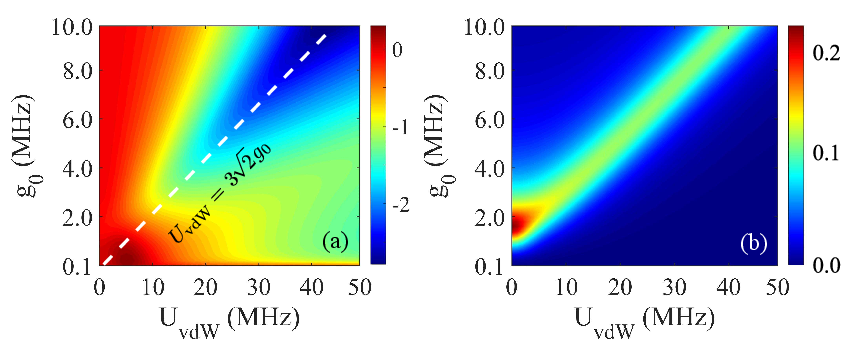}
	\caption{Plots of the second order correlation function $g^{(2)}(0)$ [panel (a)] and the mean photon number $\langle a^\dag a\rangle$ [panel (b)] as a function of the van der Waals interaction strength $U_{\rm vdW}$ and the atom-cavity coupling strength $g_0$. The dashed line in panel (a) corresponds to the condition of $U_{\rm vdW}=3\sqrt{2}g_0$, where the condition for ELA induced PB coincide with the condition for DI induced PB, leading to strong PB behavior with reasonable mean photon number.}~\label{fig:fig4}
\end{figure}
To overcome this disadvantage, we find a new condition satisfying both ELA and DI induced PBs, i.e., setting $\Delta_A=-\sqrt{2}g_0=-U_{\rm vdW}/3$. Therefore, one can obtain $U_{\rm vdW}=3\sqrt{2}g_0$, where strong single photon blockade behavior can be observed with reasonable mean photon number. Figure~\ref{fig:fig4} shows how the second order correlation function $g^{(2)}(0)$ [panel (a)] and the corresponding mean photon number [panel (b)] change as a function of $U_{\rm vdW}$ and $g_0$. Here, we set $\Delta_A=-U_{\rm vdW}/3$ and other system parameters are as same as Fig. \ref{fig:fig2}. It is clear to see that, under the condition of $U_{\rm vdW}=3\sqrt{2}g_0$, the single photon blockade behavior can be significantly improved and the corresponding second order correlation function $g^{(2)}(0)$ decreases to the value less than $10^{-2}$ as the van der Waals interaction strength $U_{\rm vdW}$ increases. At the same time, the mean photon number remains unchanged with $\langle a^\dag a\rangle\approx0.1$. The corresponding leaking rate of the cavity photons reaches up to $100$ kHz, which can be observed in experiments with current experimental techniques.

\section{Antiblockade induced emission of bunching photon pairs}
Now, let's focus on the two photon emission at the frequency of antiblockade regime, i.e., $\Delta_A\approx-U_{\rm vdW}/2$. Under this condition, as shown in Fig.~\ref{fig:fig3}(b), the state $|ee,0\rangle$ can be resonantly excited via the two photon process since the state $|ee,0\rangle$ is shifted by an anoumt of $U_{\rm vdW}$, which conpensate the probe field detuning $2\Delta_A$. As a result, the transition pathway $|gg,0\rangle\leftrightarrow|ee,0\rangle$ is dominant so that photons are strongly bunched with $g^{(2)}(0)\gg1$. Also notice that the excitation of the state $|+,1\rangle$ is still off resonant, which prevent the excitation of the state in three photon space (such as the state $|ee,1\rangle$), leading to the third order correlation function $g^{(3)}(0)<1$. Therefore, two photon blockade takes place and one can observe the emission of bunching photon pairs. 

\begin{figure}[htb]
	\includegraphics[width=\linewidth]{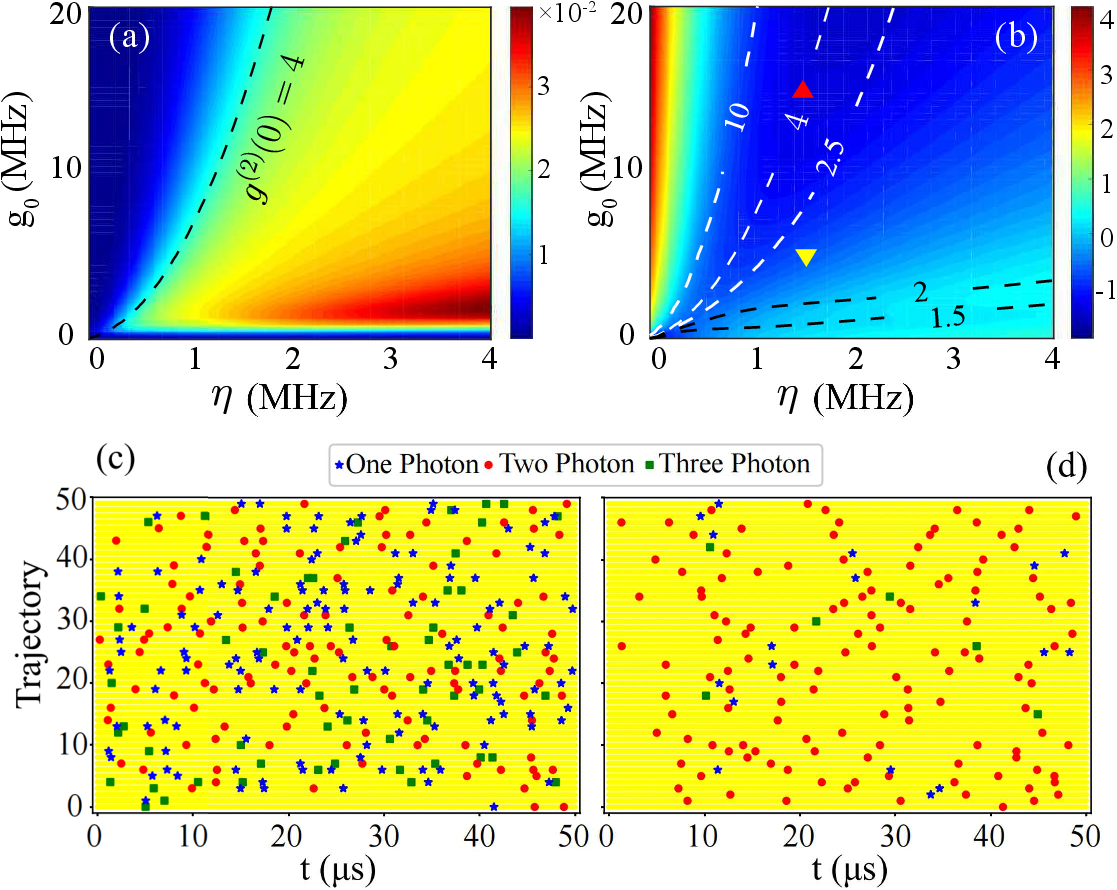}
	\caption{Plots of the mean photon number $\langle a^\dag a\rangle$ [panel (a)] and the third order correlation function $g^{(3)}(0)$ [panel (b)] as a function of the atom cavity coupling strength $g_0$ and the driving field Rabi frequency $\eta$. The dashed curves denote the values of the second order correlation function $g^{(2)}(0)$. Panels (c) and (d) show the MC simulations of photon emissions with the atom cavity coupling strength of Fig.~\ref{fig:fig5}(b) at $g_0=5$ MHz (labeled by a yellow inverted triangle) and $g_0=15$ MHz (labeled by a red triangle), respectively.} ~\label{fig:fig5}
\end{figure}
To verify the above analysis, we calculate the mean photon number $\langle a^\dag a\rangle$, the steady-state second order correlation funciton $g^{(2)}(0)$ and the third order correlation functions $g^{(3)}(0)$, respectively. In Fig.~\ref{fig:fig5}(a), the mean photon number is depicted as a function of the coupling strength $g_0$ and the driving field Rabi frequency $\eta$. The corresponding steady-state third order correlation function $g^{(3)}(0)$ is demonstrated in figure \ref{fig:fig5}(b). In panels (a) and (b), the dashed curves denote values of the second order correlation function $g^{(2)}(0)$. Here, the detuning is chosen as $\Delta_A=-U_{\rm vdW}/2$ and other system parameters are the same as those used in Fig.~\ref{fig:fig4}. It is clear to see that strong two photon blockade with $g^{(2)}(0)\gg1$ and $g^{(3)}(0)<0.1$ can be achieved by choosing suitable driving field $\eta$ and atom cavity coupling strength $g_0$. 

To gain insights into the dynamical properties of photon emission, we apply a quantum MC approach to solve the master equation with driving field Rabi frequency $\eta=1.5$.
Figures \ref{fig:fig5}(c) and (d) present a series of detection events for the atom cavity coupling strength of Fig.~\ref{fig:fig5}(b) at $g_0=5$ MHz (labeled by a yellow inverted triangle) and $g_0=15$ MHz (labeled by a red triangle), respectively. The horizontal axis represents time, and each point denotes a detection event. Although the number of total detection events in Fig.~\ref{fig:fig5}(d) is far less than that in Fig.~\ref{fig:fig5}(c) (corresponding to a smaller mean photon number), single photon and multiphoton emissions can be significantly suppressed under the strong coupling regime, i.e., $g_0=15$ MHz, leading to a purified emission of bunching photon pairs [see Fig.~\ref{fig:fig5}(d)].

\section{Conclusion}
In conclusion, we have shown how the van der Waals interaction in a two-atom cavity QED system offers a powerful knob for controlling photon statistics. By harnessing the vdW-induced shift in the two atoms excited state, we can achieve strong antibunching photon emission through destructive interference in a diamond configuration, a direct consequence of the unconventional photon blockade phenomenon. We also found that carefully adjusting the vdW interaction strength allows for an overlap between conventional and unconventional photon blockade conditions, leading to significantly enhanced single-photon emission with extremely high purity and reasonable leaking rate. Interestingly, our investigation also revealed the emission of antibunching photon pairs at the antiblockade excitation frequency, characterized by the second order correlation function $g^{(2)}(0)\gg1$. Through quantum Monte Carlo simulations, we have confirmed the feasibility of generating antibunching photon pairs by optimizing the driving field Rabi frequency and vdW interaction strength, with a purity over $90\%$ (the third order correlation function $g^{(3)}(0)<0.1$). These capabilities have significant implications for quantum technologies, such as quantum light sources, quantum key distribution and optical quantum computing. Meanwhile, the controlled generation of antibunching photon pairs opens avenues for advancements in quantum metrology and quantum imaging, offering enhanced sensitivity and resolution. 

\vskip 5pt
\begin{acknowledgments}
CJZ and YPY thank the support from the National Nature Science Foundation (Grant No. 11774262, No. 12274326 and No. 12174288, No. 12274131) and the National
Key R\&D Program of China (Grant No. 2021YFA1400602). CJZ acknowledges Prof. G. S. Agarwal's inspiration at Texas A\&M University
%
\end{acknowledgments}

\bibliography{ref}

\begin{thebibliography}{79}%
\makeatletter
\providecommand \@ifxundefined [1]{%
 \@ifx{#1\undefined}
}%
\providecommand \@ifnum [1]{%
 \ifnum #1\expandafter \@firstoftwo
 \else \expandafter \@secondoftwo
 \fi
}%
\providecommand \@ifx [1]{%
 \ifx #1\expandafter \@firstoftwo
 \else \expandafter \@secondoftwo
 \fi
}%
\providecommand \natexlab [1]{#1}%
\providecommand \enquote  [1]{``#1''}%
\providecommand \bibnamefont  [1]{#1}%
\providecommand \bibfnamefont [1]{#1}%
\providecommand \citenamefont [1]{#1}%
\providecommand \href@noop [0]{\@secondoftwo}%
\providecommand \href [0]{\begingroup \@sanitize@url \@href}%
\providecommand \@href[1]{\@@startlink{#1}\@@href}%
\providecommand \@@href[1]{\endgroup#1\@@endlink}%
\providecommand \@sanitize@url [0]{\catcode `\\12\catcode `\$12\catcode
  `\&12\catcode `\#12\catcode `\^12\catcode `\_12\catcode `\%12\relax}%
\providecommand \@@startlink[1]{}%
\providecommand \@@endlink[0]{}%
\providecommand \url  [0]{\begingroup\@sanitize@url \@url }%
\providecommand \@url [1]{\endgroup\@href {#1}{\urlprefix }}%
\providecommand \urlprefix  [0]{URL }%
\providecommand \Eprint [0]{\href }%
\providecommand \doibase [0]{https://doi.org/}%
\providecommand \selectlanguage [0]{\@gobble}%
\providecommand \bibinfo  [0]{\@secondoftwo}%
\providecommand \bibfield  [0]{\@secondoftwo}%
\providecommand \translation [1]{[#1]}%
\providecommand \BibitemOpen [0]{}%
\providecommand \bibitemStop [0]{}%
\providecommand \bibitemNoStop [0]{.\EOS\space}%
\providecommand \EOS [0]{\spacefactor3000\relax}%
\providecommand \BibitemShut  [1]{\csname bibitem#1\endcsname}%
\let\auto@bib@innerbib\@empty
\bibitem [{\citenamefont {Zou}\ and\ \citenamefont
  {Mandel}(1990)}]{zou1990photon}%
  \BibitemOpen
  \bibfield  {author} {\bibinfo {author} {\bibfnamefont {X.}~\bibnamefont
  {Zou}}\ and\ \bibinfo {author} {\bibfnamefont {L.}~\bibnamefont {Mandel}},\
  }\bibfield  {title} {\bibinfo {title} {Photon-antibunching and sub-poissonian
  photon statistics},\ }\href@noop {} {\bibfield  {journal} {\bibinfo
  {journal} {Physical Review A}\ }\textbf {\bibinfo {volume} {41}},\ \bibinfo
  {pages} {475} (\bibinfo {year} {1990})}\BibitemShut {NoStop}%
\bibitem [{\citenamefont {Lemonde}\ \emph {et~al.}(2014)\citenamefont
  {Lemonde}, \citenamefont {Didier},\ and\ \citenamefont
  {Clerk}}]{lemonde2014antibunching}%
  \BibitemOpen
  \bibfield  {author} {\bibinfo {author} {\bibfnamefont {M.-A.}\ \bibnamefont
  {Lemonde}}, \bibinfo {author} {\bibfnamefont {N.}~\bibnamefont {Didier}},\
  and\ \bibinfo {author} {\bibfnamefont {A.~A.}\ \bibnamefont {Clerk}},\
  }\bibfield  {title} {\bibinfo {title} {Antibunching and unconventional photon
  blockade with gaussian squeezed states},\ }\href@noop {} {\bibfield
  {journal} {\bibinfo  {journal} {Physical Review A}\ }\textbf {\bibinfo
  {volume} {90}},\ \bibinfo {pages} {063824} (\bibinfo {year}
  {2014})}\BibitemShut {NoStop}%
\bibitem [{\citenamefont {Ren}\ \emph {et~al.}(2021)\citenamefont {Ren},
  \citenamefont {Duan}, \citenamefont {Xie}, \citenamefont {Shao},\ and\
  \citenamefont {Duan}}]{ren2021antibunched}%
  \BibitemOpen
  \bibfield  {author} {\bibinfo {author} {\bibfnamefont {Y.}~\bibnamefont
  {Ren}}, \bibinfo {author} {\bibfnamefont {S.}~\bibnamefont {Duan}}, \bibinfo
  {author} {\bibfnamefont {W.}~\bibnamefont {Xie}}, \bibinfo {author}
  {\bibfnamefont {Y.}~\bibnamefont {Shao}},\ and\ \bibinfo {author}
  {\bibfnamefont {Z.}~\bibnamefont {Duan}},\ }\bibfield  {title} {\bibinfo
  {title} {Antibunched photon-pair source based on photon blockade in a
  nondegenerate optical parametric oscillator},\ }\href@noop {} {\bibfield
  {journal} {\bibinfo  {journal} {Physical Review A}\ }\textbf {\bibinfo
  {volume} {103}},\ \bibinfo {pages} {053710} (\bibinfo {year}
  {2021})}\BibitemShut {NoStop}%
\bibitem [{\citenamefont {Chen}\ \emph {et~al.}(2022)\citenamefont {Chen},
  \citenamefont {Tang}, \citenamefont {Tang}, \citenamefont {Wu},\ and\
  \citenamefont {Xia}}]{chen2022photon}%
  \BibitemOpen
  \bibfield  {author} {\bibinfo {author} {\bibfnamefont {M.}~\bibnamefont
  {Chen}}, \bibinfo {author} {\bibfnamefont {J.}~\bibnamefont {Tang}}, \bibinfo
  {author} {\bibfnamefont {L.}~\bibnamefont {Tang}}, \bibinfo {author}
  {\bibfnamefont {H.}~\bibnamefont {Wu}},\ and\ \bibinfo {author}
  {\bibfnamefont {K.}~\bibnamefont {Xia}},\ }\bibfield  {title} {\bibinfo
  {title} {Photon blockade and single-photon generation with multiple quantum
  emitters},\ }\href@noop {} {\bibfield  {journal} {\bibinfo  {journal}
  {Physical Review Research}\ }\textbf {\bibinfo {volume} {4}},\ \bibinfo
  {pages} {033083} (\bibinfo {year} {2022})}\BibitemShut {NoStop}%
\bibitem [{\citenamefont {Lu}\ \emph {et~al.}(2021)\citenamefont {Lu},
  \citenamefont {Liu}, \citenamefont {Liao},\ and\ \citenamefont
  {Wang}}]{lu2021plasmonic}%
  \BibitemOpen
  \bibfield  {author} {\bibinfo {author} {\bibfnamefont {Y.-W.}\ \bibnamefont
  {Lu}}, \bibinfo {author} {\bibfnamefont {J.-F.}\ \bibnamefont {Liu}},
  \bibinfo {author} {\bibfnamefont {Z.}~\bibnamefont {Liao}},\ and\ \bibinfo
  {author} {\bibfnamefont {X.-H.}\ \bibnamefont {Wang}},\ }\bibfield  {title}
  {\bibinfo {title} {Plasmonic-photonic cavity for high-efficiency
  single-photon blockade},\ }\href@noop {} {\bibfield  {journal} {\bibinfo
  {journal} {Science China Physics, Mechanics \& Astronomy}\ }\textbf {\bibinfo
  {volume} {64}},\ \bibinfo {pages} {274212} (\bibinfo {year}
  {2021})}\BibitemShut {NoStop}%
\bibitem [{\citenamefont {Scully}\ and\ \citenamefont
  {Zubairy}(1999)}]{scully1999quantum}%
  \BibitemOpen
  \bibfield  {author} {\bibinfo {author} {\bibfnamefont {M.~O.}\ \bibnamefont
  {Scully}}\ and\ \bibinfo {author} {\bibfnamefont {M.~S.}\ \bibnamefont
  {Zubairy}},\ }\href@noop {} {\bibinfo {title} {Quantum optics}} (\bibinfo
  {year} {1999})\BibitemShut {NoStop}%
\bibitem [{\citenamefont {Birnbaum}\ \emph {et~al.}(2005)\citenamefont
  {Birnbaum}, \citenamefont {Boca}, \citenamefont {Miller}, \citenamefont
  {Boozer}, \citenamefont {Northup},\ and\ \citenamefont
  {Kimble}}]{birnbaum2005photon}%
  \BibitemOpen
  \bibfield  {author} {\bibinfo {author} {\bibfnamefont {K.~M.}\ \bibnamefont
  {Birnbaum}}, \bibinfo {author} {\bibfnamefont {A.}~\bibnamefont {Boca}},
  \bibinfo {author} {\bibfnamefont {R.}~\bibnamefont {Miller}}, \bibinfo
  {author} {\bibfnamefont {A.~D.}\ \bibnamefont {Boozer}}, \bibinfo {author}
  {\bibfnamefont {T.~E.}\ \bibnamefont {Northup}},\ and\ \bibinfo {author}
  {\bibfnamefont {H.~J.}\ \bibnamefont {Kimble}},\ }\bibfield  {title}
  {\bibinfo {title} {Photon blockade in an optical cavity with one trapped
  atom},\ }\href@noop {} {\bibfield  {journal} {\bibinfo  {journal} {Nature}\
  }\textbf {\bibinfo {volume} {436}},\ \bibinfo {pages} {87} (\bibinfo {year}
  {2005})}\BibitemShut {NoStop}%
\bibitem [{\citenamefont {Hacker}\ \emph {et~al.}(2016)\citenamefont {Hacker},
  \citenamefont {Welte}, \citenamefont {Rempe},\ and\ \citenamefont
  {Ritter}}]{hacker2016photon}%
  \BibitemOpen
  \bibfield  {author} {\bibinfo {author} {\bibfnamefont {B.}~\bibnamefont
  {Hacker}}, \bibinfo {author} {\bibfnamefont {S.}~\bibnamefont {Welte}},
  \bibinfo {author} {\bibfnamefont {G.}~\bibnamefont {Rempe}},\ and\ \bibinfo
  {author} {\bibfnamefont {S.}~\bibnamefont {Ritter}},\ }\bibfield  {title}
  {\bibinfo {title} {A photon--photon quantum gate based on a single atom in an
  optical resonator},\ }\href@noop {} {\bibfield  {journal} {\bibinfo
  {journal} {Nature}\ }\textbf {\bibinfo {volume} {536}},\ \bibinfo {pages}
  {193} (\bibinfo {year} {2016})}\BibitemShut {NoStop}%
\bibitem [{\citenamefont {Tiarks}\ \emph {et~al.}(2019)\citenamefont {Tiarks},
  \citenamefont {Schmidt-Eberle}, \citenamefont {Stolz}, \citenamefont
  {Rempe},\ and\ \citenamefont {D{\"u}rr}}]{tiarks2019photon}%
  \BibitemOpen
  \bibfield  {author} {\bibinfo {author} {\bibfnamefont {D.}~\bibnamefont
  {Tiarks}}, \bibinfo {author} {\bibfnamefont {S.}~\bibnamefont
  {Schmidt-Eberle}}, \bibinfo {author} {\bibfnamefont {T.}~\bibnamefont
  {Stolz}}, \bibinfo {author} {\bibfnamefont {G.}~\bibnamefont {Rempe}},\ and\
  \bibinfo {author} {\bibfnamefont {S.}~\bibnamefont {D{\"u}rr}},\ }\bibfield
  {title} {\bibinfo {title} {A photon--photon quantum gate based on rydberg
  interactions},\ }\href@noop {} {\bibfield  {journal} {\bibinfo  {journal}
  {Nature Physics}\ }\textbf {\bibinfo {volume} {15}},\ \bibinfo {pages} {124}
  (\bibinfo {year} {2019})}\BibitemShut {NoStop}%
\bibitem [{\citenamefont {Li}\ \emph {et~al.}(2020)\citenamefont {Li},
  \citenamefont {Zhang}, \citenamefont {Tang}, \citenamefont {Dong},
  \citenamefont {Guo},\ and\ \citenamefont {Zou}}]{li2020photon}%
  \BibitemOpen
  \bibfield  {author} {\bibinfo {author} {\bibfnamefont {M.}~\bibnamefont
  {Li}}, \bibinfo {author} {\bibfnamefont {Y.-L.}\ \bibnamefont {Zhang}},
  \bibinfo {author} {\bibfnamefont {H.~X.}\ \bibnamefont {Tang}}, \bibinfo
  {author} {\bibfnamefont {C.-H.}\ \bibnamefont {Dong}}, \bibinfo {author}
  {\bibfnamefont {G.-C.}\ \bibnamefont {Guo}},\ and\ \bibinfo {author}
  {\bibfnamefont {C.-L.}\ \bibnamefont {Zou}},\ }\bibfield  {title} {\bibinfo
  {title} {Photon-photon quantum phase gate in a photonic molecule with $\chi$
  (2) nonlinearity},\ }\href@noop {} {\bibfield  {journal} {\bibinfo  {journal}
  {Physical Review Applied}\ }\textbf {\bibinfo {volume} {13}},\ \bibinfo
  {pages} {044013} (\bibinfo {year} {2020})}\BibitemShut {NoStop}%
\bibitem [{\citenamefont {Dayan}\ \emph {et~al.}(2008)\citenamefont {Dayan},
  \citenamefont {Parkins}, \citenamefont {Aoki}, \citenamefont {Ostby},
  \citenamefont {Vahala},\ and\ \citenamefont {Kimble}}]{dayan2008photon}%
  \BibitemOpen
  \bibfield  {author} {\bibinfo {author} {\bibfnamefont {B.}~\bibnamefont
  {Dayan}}, \bibinfo {author} {\bibfnamefont {A.}~\bibnamefont {Parkins}},
  \bibinfo {author} {\bibfnamefont {T.}~\bibnamefont {Aoki}}, \bibinfo {author}
  {\bibfnamefont {E.}~\bibnamefont {Ostby}}, \bibinfo {author} {\bibfnamefont
  {K.}~\bibnamefont {Vahala}},\ and\ \bibinfo {author} {\bibfnamefont
  {H.}~\bibnamefont {Kimble}},\ }\bibfield  {title} {\bibinfo {title} {A photon
  turnstile dynamically regulated by one atom},\ }\href@noop {} {\bibfield
  {journal} {\bibinfo  {journal} {Science}\ }\textbf {\bibinfo {volume}
  {319}},\ \bibinfo {pages} {1062} (\bibinfo {year} {2008})}\BibitemShut
  {NoStop}%
\bibitem [{\citenamefont {Huang}\ \emph {et~al.}(2018)\citenamefont {Huang},
  \citenamefont {Miranowicz}, \citenamefont {Liao}, \citenamefont {Nori},\ and\
  \citenamefont {Jing}}]{huang2018nonreciprocal}%
  \BibitemOpen
  \bibfield  {author} {\bibinfo {author} {\bibfnamefont {R.}~\bibnamefont
  {Huang}}, \bibinfo {author} {\bibfnamefont {A.}~\bibnamefont {Miranowicz}},
  \bibinfo {author} {\bibfnamefont {J.-Q.}\ \bibnamefont {Liao}}, \bibinfo
  {author} {\bibfnamefont {F.}~\bibnamefont {Nori}},\ and\ \bibinfo {author}
  {\bibfnamefont {H.}~\bibnamefont {Jing}},\ }\bibfield  {title} {\bibinfo
  {title} {Nonreciprocal photon blockade},\ }\href@noop {} {\bibfield
  {journal} {\bibinfo  {journal} {Physical review letters}\ }\textbf {\bibinfo
  {volume} {121}},\ \bibinfo {pages} {153601} (\bibinfo {year}
  {2018})}\BibitemShut {NoStop}%
\bibitem [{\citenamefont {M{\"u}ller}\ \emph {et~al.}(2015)\citenamefont
  {M{\"u}ller}, \citenamefont {Rundquist}, \citenamefont {Fischer},
  \citenamefont {Sarmiento}, \citenamefont {Lagoudakis}, \citenamefont
  {Kelaita}, \citenamefont {Munoz}, \citenamefont {Del~Valle}, \citenamefont
  {Laussy},\ and\ \citenamefont {Vu{\v{c}}kovi{\'c}}}]{muller2015coherent}%
  \BibitemOpen
  \bibfield  {author} {\bibinfo {author} {\bibfnamefont {K.}~\bibnamefont
  {M{\"u}ller}}, \bibinfo {author} {\bibfnamefont {A.}~\bibnamefont
  {Rundquist}}, \bibinfo {author} {\bibfnamefont {K.~A.}\ \bibnamefont
  {Fischer}}, \bibinfo {author} {\bibfnamefont {T.}~\bibnamefont {Sarmiento}},
  \bibinfo {author} {\bibfnamefont {K.~G.}\ \bibnamefont {Lagoudakis}},
  \bibinfo {author} {\bibfnamefont {Y.~A.}\ \bibnamefont {Kelaita}}, \bibinfo
  {author} {\bibfnamefont {C.~S.}\ \bibnamefont {Munoz}}, \bibinfo {author}
  {\bibfnamefont {E.}~\bibnamefont {Del~Valle}}, \bibinfo {author}
  {\bibfnamefont {F.~P.}\ \bibnamefont {Laussy}},\ and\ \bibinfo {author}
  {\bibfnamefont {J.}~\bibnamefont {Vu{\v{c}}kovi{\'c}}},\ }\bibfield  {title}
  {\bibinfo {title} {Coherent generation of nonclassical light on chip via
  detuned photon blockade},\ }\href@noop {} {\bibfield  {journal} {\bibinfo
  {journal} {Physical review letters}\ }\textbf {\bibinfo {volume} {114}},\
  \bibinfo {pages} {233601} (\bibinfo {year} {2015})}\BibitemShut {NoStop}%
\bibitem [{\citenamefont {Faraon}\ \emph {et~al.}(2010)\citenamefont {Faraon},
  \citenamefont {Majumdar},\ and\ \citenamefont
  {Vu{\v{c}}kovi{\'c}}}]{faraon2010generation}%
  \BibitemOpen
  \bibfield  {author} {\bibinfo {author} {\bibfnamefont {A.}~\bibnamefont
  {Faraon}}, \bibinfo {author} {\bibfnamefont {A.}~\bibnamefont {Majumdar}},\
  and\ \bibinfo {author} {\bibfnamefont {J.}~\bibnamefont
  {Vu{\v{c}}kovi{\'c}}},\ }\bibfield  {title} {\bibinfo {title} {Generation of
  nonclassical states of light via photon blockade in optical nanocavities},\
  }\href@noop {} {\bibfield  {journal} {\bibinfo  {journal} {Physical Review
  A}\ }\textbf {\bibinfo {volume} {81}},\ \bibinfo {pages} {033838} (\bibinfo
  {year} {2010})}\BibitemShut {NoStop}%
\bibitem [{\citenamefont {Faraon}\ \emph {et~al.}(2008)\citenamefont {Faraon},
  \citenamefont {Fushman}, \citenamefont {Englund}, \citenamefont {Stoltz},
  \citenamefont {Petroff},\ and\ \citenamefont
  {Vu{\v{c}}kovi{\'c}}}]{faraon2008coherent}%
  \BibitemOpen
  \bibfield  {author} {\bibinfo {author} {\bibfnamefont {A.}~\bibnamefont
  {Faraon}}, \bibinfo {author} {\bibfnamefont {I.}~\bibnamefont {Fushman}},
  \bibinfo {author} {\bibfnamefont {D.}~\bibnamefont {Englund}}, \bibinfo
  {author} {\bibfnamefont {N.}~\bibnamefont {Stoltz}}, \bibinfo {author}
  {\bibfnamefont {P.}~\bibnamefont {Petroff}},\ and\ \bibinfo {author}
  {\bibfnamefont {J.}~\bibnamefont {Vu{\v{c}}kovi{\'c}}},\ }\bibfield  {title}
  {\bibinfo {title} {Coherent generation of non-classical light on a chip via
  photon-induced tunnelling and blockade},\ }\href@noop {} {\bibfield
  {journal} {\bibinfo  {journal} {Nature Physics}\ }\textbf {\bibinfo {volume}
  {4}},\ \bibinfo {pages} {859} (\bibinfo {year} {2008})}\BibitemShut {NoStop}%
\bibitem [{\citenamefont {Flayac}\ \emph {et~al.}(2015)\citenamefont {Flayac},
  \citenamefont {Gerace},\ and\ \citenamefont {Savona}}]{flayac2015all}%
  \BibitemOpen
  \bibfield  {author} {\bibinfo {author} {\bibfnamefont {H.}~\bibnamefont
  {Flayac}}, \bibinfo {author} {\bibfnamefont {D.}~\bibnamefont {Gerace}},\
  and\ \bibinfo {author} {\bibfnamefont {V.}~\bibnamefont {Savona}},\
  }\bibfield  {title} {\bibinfo {title} {An all-silicon single-photon source by
  unconventional photon blockade},\ }\href@noop {} {\bibfield  {journal}
  {\bibinfo  {journal} {Scientific reports}\ }\textbf {\bibinfo {volume} {5}},\
  \bibinfo {pages} {11223} (\bibinfo {year} {2015})}\BibitemShut {NoStop}%
\bibitem [{\citenamefont {Tang}\ \emph {et~al.}(2021)\citenamefont {Tang},
  \citenamefont {Tang}, \citenamefont {Wu}, \citenamefont {Wu}, \citenamefont
  {Sun}, \citenamefont {Zhang}, \citenamefont {Li}, \citenamefont {Lu},
  \citenamefont {Xiao}, \citenamefont {Xia} \emph {et~al.}}]{tang2021towards}%
  \BibitemOpen
  \bibfield  {author} {\bibinfo {author} {\bibfnamefont {J.}~\bibnamefont
  {Tang}}, \bibinfo {author} {\bibfnamefont {L.}~\bibnamefont {Tang}}, \bibinfo
  {author} {\bibfnamefont {H.}~\bibnamefont {Wu}}, \bibinfo {author}
  {\bibfnamefont {Y.}~\bibnamefont {Wu}}, \bibinfo {author} {\bibfnamefont
  {H.}~\bibnamefont {Sun}}, \bibinfo {author} {\bibfnamefont {H.}~\bibnamefont
  {Zhang}}, \bibinfo {author} {\bibfnamefont {T.}~\bibnamefont {Li}}, \bibinfo
  {author} {\bibfnamefont {Y.}~\bibnamefont {Lu}}, \bibinfo {author}
  {\bibfnamefont {M.}~\bibnamefont {Xiao}}, \bibinfo {author} {\bibfnamefont
  {K.}~\bibnamefont {Xia}}, \emph {et~al.},\ }\bibfield  {title} {\bibinfo
  {title} {Towards on-demand heralded single-photon sources via photon
  blockade},\ }\href@noop {} {\bibfield  {journal} {\bibinfo  {journal}
  {Physical Review Applied}\ }\textbf {\bibinfo {volume} {15}},\ \bibinfo
  {pages} {064020} (\bibinfo {year} {2021})}\BibitemShut {NoStop}%
\bibitem [{\citenamefont {Majumdar}\ and\ \citenamefont
  {Gerace}(2013)}]{majumdar2013single}%
  \BibitemOpen
  \bibfield  {author} {\bibinfo {author} {\bibfnamefont {A.}~\bibnamefont
  {Majumdar}}\ and\ \bibinfo {author} {\bibfnamefont {D.}~\bibnamefont
  {Gerace}},\ }\bibfield  {title} {\bibinfo {title} {Single-photon blockade in
  doubly resonant nanocavities with second-order nonlinearity},\ }\href@noop {}
  {\bibfield  {journal} {\bibinfo  {journal} {Physical Review B}\ }\textbf
  {\bibinfo {volume} {87}},\ \bibinfo {pages} {235319} (\bibinfo {year}
  {2013})}\BibitemShut {NoStop}%
\bibitem [{\citenamefont {Liew}\ and\ \citenamefont
  {Savona}(2010)}]{liew2010single}%
  \BibitemOpen
  \bibfield  {author} {\bibinfo {author} {\bibfnamefont {T.~C.}\ \bibnamefont
  {Liew}}\ and\ \bibinfo {author} {\bibfnamefont {V.}~\bibnamefont {Savona}},\
  }\bibfield  {title} {\bibinfo {title} {Single photons from coupled quantum
  modes},\ }\href@noop {} {\bibfield  {journal} {\bibinfo  {journal} {Physical
  Review Letters}\ }\textbf {\bibinfo {volume} {104}},\ \bibinfo {pages}
  {183601} (\bibinfo {year} {2010})}\BibitemShut {NoStop}%
\bibitem [{\citenamefont {Flayac}\ and\ \citenamefont
  {Savona}(2017)}]{flayac2017unconventional}%
  \BibitemOpen
  \bibfield  {author} {\bibinfo {author} {\bibfnamefont {H.}~\bibnamefont
  {Flayac}}\ and\ \bibinfo {author} {\bibfnamefont {V.}~\bibnamefont
  {Savona}},\ }\bibfield  {title} {\bibinfo {title} {Unconventional photon
  blockade},\ }\href@noop {} {\bibfield  {journal} {\bibinfo  {journal}
  {Physical Review A}\ }\textbf {\bibinfo {volume} {96}},\ \bibinfo {pages}
  {053810} (\bibinfo {year} {2017})}\BibitemShut {NoStop}%
\bibitem [{\citenamefont {Zubizarreta~Casalengua}\ \emph
  {et~al.}(2020)\citenamefont {Zubizarreta~Casalengua}, \citenamefont
  {L{\'o}pez~Carre{\~n}o}, \citenamefont {Laussy},\ and\ \citenamefont
  {Valle}}]{zubizarreta2020conventional}%
  \BibitemOpen
  \bibfield  {author} {\bibinfo {author} {\bibfnamefont {E.}~\bibnamefont
  {Zubizarreta~Casalengua}}, \bibinfo {author} {\bibfnamefont {J.~C.}\
  \bibnamefont {L{\'o}pez~Carre{\~n}o}}, \bibinfo {author} {\bibfnamefont
  {F.~P.}\ \bibnamefont {Laussy}},\ and\ \bibinfo {author} {\bibfnamefont
  {E.~d.}\ \bibnamefont {Valle}},\ }\bibfield  {title} {\bibinfo {title}
  {Conventional and unconventional photon statistics},\ }\href@noop {}
  {\bibfield  {journal} {\bibinfo  {journal} {Laser \& Photonics Reviews}\
  }\textbf {\bibinfo {volume} {14}},\ \bibinfo {pages} {1900279} (\bibinfo
  {year} {2020})}\BibitemShut {NoStop}%
\bibitem [{\citenamefont {Bamba}\ \emph {et~al.}(2011)\citenamefont {Bamba},
  \citenamefont {Imamo{\u{g}}lu}, \citenamefont {Carusotto},\ and\
  \citenamefont {Ciuti}}]{bamba2011origin}%
  \BibitemOpen
  \bibfield  {author} {\bibinfo {author} {\bibfnamefont {M.}~\bibnamefont
  {Bamba}}, \bibinfo {author} {\bibfnamefont {A.}~\bibnamefont
  {Imamo{\u{g}}lu}}, \bibinfo {author} {\bibfnamefont {I.}~\bibnamefont
  {Carusotto}},\ and\ \bibinfo {author} {\bibfnamefont {C.}~\bibnamefont
  {Ciuti}},\ }\bibfield  {title} {\bibinfo {title} {Origin of strong photon
  antibunching in weakly nonlinear photonic molecules},\ }\href@noop {}
  {\bibfield  {journal} {\bibinfo  {journal} {Physical Review A}\ }\textbf
  {\bibinfo {volume} {83}},\ \bibinfo {pages} {021802} (\bibinfo {year}
  {2011})}\BibitemShut {NoStop}%
\bibitem [{\citenamefont {Hou}\ \emph {et~al.}(2019)\citenamefont {Hou},
  \citenamefont {Zhu}, \citenamefont {Yang},\ and\ \citenamefont
  {Agarwal}}]{hou2019interfering}%
  \BibitemOpen
  \bibfield  {author} {\bibinfo {author} {\bibfnamefont {K.}~\bibnamefont
  {Hou}}, \bibinfo {author} {\bibfnamefont {C.}~\bibnamefont {Zhu}}, \bibinfo
  {author} {\bibfnamefont {Y.}~\bibnamefont {Yang}},\ and\ \bibinfo {author}
  {\bibfnamefont {G.}~\bibnamefont {Agarwal}},\ }\bibfield  {title} {\bibinfo
  {title} {Interfering pathways for photon blockade in cavity qed with one and
  two qubits},\ }\href@noop {} {\bibfield  {journal} {\bibinfo  {journal}
  {Physical Review A}\ }\textbf {\bibinfo {volume} {100}},\ \bibinfo {pages}
  {063817} (\bibinfo {year} {2019})}\BibitemShut {NoStop}%
\bibitem [{\citenamefont {Zhou}\ \emph {et~al.}(2025)\citenamefont {Zhou},
  \citenamefont {Liu}, \citenamefont {Su}, \citenamefont {Zhang}, \citenamefont
  {Wu}, \citenamefont {Chen}, \citenamefont {Shi}, \citenamefont {Shen},\ and\
  \citenamefont {Yang}}]{zhou2025universal}%
  \BibitemOpen
  \bibfield  {author} {\bibinfo {author} {\bibfnamefont {Y.-H.}\ \bibnamefont
  {Zhou}}, \bibinfo {author} {\bibfnamefont {T.}~\bibnamefont {Liu}}, \bibinfo
  {author} {\bibfnamefont {Q.-P.}\ \bibnamefont {Su}}, \bibinfo {author}
  {\bibfnamefont {X.-Y.}\ \bibnamefont {Zhang}}, \bibinfo {author}
  {\bibfnamefont {Q.-C.}\ \bibnamefont {Wu}}, \bibinfo {author} {\bibfnamefont
  {D.-X.}\ \bibnamefont {Chen}}, \bibinfo {author} {\bibfnamefont {Z.-C.}\
  \bibnamefont {Shi}}, \bibinfo {author} {\bibfnamefont {H.}~\bibnamefont
  {Shen}},\ and\ \bibinfo {author} {\bibfnamefont {C.-P.}\ \bibnamefont
  {Yang}},\ }\bibfield  {title} {\bibinfo {title} {Universal photon blockade},\
  }\href@noop {} {\bibfield  {journal} {\bibinfo  {journal} {Physical Review
  Letters}\ }\textbf {\bibinfo {volume} {134}},\ \bibinfo {pages} {183601}
  (\bibinfo {year} {2025})}\BibitemShut {NoStop}%
\bibitem [{\citenamefont {Snijders}\ \emph {et~al.}(2018)\citenamefont
  {Snijders}, \citenamefont {Frey}, \citenamefont {Norman}, \citenamefont
  {Flayac}, \citenamefont {Savona}, \citenamefont {Gossard}, \citenamefont
  {Bowers}, \citenamefont {Van~Exter}, \citenamefont {Bouwmeester},\ and\
  \citenamefont {L{\"o}ffler}}]{snijders2018observation}%
  \BibitemOpen
  \bibfield  {author} {\bibinfo {author} {\bibfnamefont {H.}~\bibnamefont
  {Snijders}}, \bibinfo {author} {\bibfnamefont {J.}~\bibnamefont {Frey}},
  \bibinfo {author} {\bibfnamefont {J.}~\bibnamefont {Norman}}, \bibinfo
  {author} {\bibfnamefont {H.}~\bibnamefont {Flayac}}, \bibinfo {author}
  {\bibfnamefont {V.}~\bibnamefont {Savona}}, \bibinfo {author} {\bibfnamefont
  {A.}~\bibnamefont {Gossard}}, \bibinfo {author} {\bibfnamefont
  {J.}~\bibnamefont {Bowers}}, \bibinfo {author} {\bibfnamefont
  {M.}~\bibnamefont {Van~Exter}}, \bibinfo {author} {\bibfnamefont
  {D.}~\bibnamefont {Bouwmeester}},\ and\ \bibinfo {author} {\bibfnamefont
  {W.}~\bibnamefont {L{\"o}ffler}},\ }\bibfield  {title} {\bibinfo {title}
  {Observation of the unconventional photon blockade},\ }\href@noop {}
  {\bibfield  {journal} {\bibinfo  {journal} {Physical review letters}\
  }\textbf {\bibinfo {volume} {121}},\ \bibinfo {pages} {043601} (\bibinfo
  {year} {2018})}\BibitemShut {NoStop}%
\bibitem [{\citenamefont {Ridolfo}\ \emph {et~al.}(2012)\citenamefont
  {Ridolfo}, \citenamefont {Leib}, \citenamefont {Savasta},\ and\ \citenamefont
  {Hartmann}}]{ridolfo2012photon}%
  \BibitemOpen
  \bibfield  {author} {\bibinfo {author} {\bibfnamefont {A.}~\bibnamefont
  {Ridolfo}}, \bibinfo {author} {\bibfnamefont {M.}~\bibnamefont {Leib}},
  \bibinfo {author} {\bibfnamefont {S.}~\bibnamefont {Savasta}},\ and\ \bibinfo
  {author} {\bibfnamefont {M.~J.}\ \bibnamefont {Hartmann}},\ }\bibfield
  {title} {\bibinfo {title} {Photon blockade in the ultrastrong coupling
  regime},\ }\href@noop {} {\bibfield  {journal} {\bibinfo  {journal} {Physical
  review letters}\ }\textbf {\bibinfo {volume} {109}},\ \bibinfo {pages}
  {193602} (\bibinfo {year} {2012})}\BibitemShut {NoStop}%
\bibitem [{\citenamefont {Peyronel}\ \emph {et~al.}(2012)\citenamefont
  {Peyronel}, \citenamefont {Firstenberg}, \citenamefont {Liang}, \citenamefont
  {Hofferberth}, \citenamefont {Gorshkov}, \citenamefont {Pohl}, \citenamefont
  {Lukin},\ and\ \citenamefont {Vuleti{\'c}}}]{peyronel2012quantum}%
  \BibitemOpen
  \bibfield  {author} {\bibinfo {author} {\bibfnamefont {T.}~\bibnamefont
  {Peyronel}}, \bibinfo {author} {\bibfnamefont {O.}~\bibnamefont
  {Firstenberg}}, \bibinfo {author} {\bibfnamefont {Q.-Y.}\ \bibnamefont
  {Liang}}, \bibinfo {author} {\bibfnamefont {S.}~\bibnamefont {Hofferberth}},
  \bibinfo {author} {\bibfnamefont {A.~V.}\ \bibnamefont {Gorshkov}}, \bibinfo
  {author} {\bibfnamefont {T.}~\bibnamefont {Pohl}}, \bibinfo {author}
  {\bibfnamefont {M.~D.}\ \bibnamefont {Lukin}},\ and\ \bibinfo {author}
  {\bibfnamefont {V.}~\bibnamefont {Vuleti{\'c}}},\ }\bibfield  {title}
  {\bibinfo {title} {Quantum nonlinear optics with single photons enabled by
  strongly interacting atoms},\ }\href@noop {} {\bibfield  {journal} {\bibinfo
  {journal} {Nature}\ }\textbf {\bibinfo {volume} {488}},\ \bibinfo {pages}
  {57} (\bibinfo {year} {2012})}\BibitemShut {NoStop}%
\bibitem [{\citenamefont {Radulaski}\ \emph {et~al.}(2017)\citenamefont
  {Radulaski}, \citenamefont {Fischer}, \citenamefont {Lagoudakis},
  \citenamefont {Zhang},\ and\ \citenamefont
  {Vu{\v{c}}kovi{\'c}}}]{radulaski2017photon}%
  \BibitemOpen
  \bibfield  {author} {\bibinfo {author} {\bibfnamefont {M.}~\bibnamefont
  {Radulaski}}, \bibinfo {author} {\bibfnamefont {K.~A.}\ \bibnamefont
  {Fischer}}, \bibinfo {author} {\bibfnamefont {K.~G.}\ \bibnamefont
  {Lagoudakis}}, \bibinfo {author} {\bibfnamefont {J.~L.}\ \bibnamefont
  {Zhang}},\ and\ \bibinfo {author} {\bibfnamefont {J.}~\bibnamefont
  {Vu{\v{c}}kovi{\'c}}},\ }\bibfield  {title} {\bibinfo {title} {Photon
  blockade in two-emitter-cavity systems},\ }\href@noop {} {\bibfield
  {journal} {\bibinfo  {journal} {Physical Review A}\ }\textbf {\bibinfo
  {volume} {96}},\ \bibinfo {pages} {011801} (\bibinfo {year}
  {2017})}\BibitemShut {NoStop}%
\bibitem [{\citenamefont {Trivedi}\ \emph {et~al.}(2019)\citenamefont
  {Trivedi}, \citenamefont {Radulaski}, \citenamefont {Fischer}, \citenamefont
  {Fan},\ and\ \citenamefont {Vu{\v{c}}kovi{\'c}}}]{trivedi2019photon}%
  \BibitemOpen
  \bibfield  {author} {\bibinfo {author} {\bibfnamefont {R.}~\bibnamefont
  {Trivedi}}, \bibinfo {author} {\bibfnamefont {M.}~\bibnamefont {Radulaski}},
  \bibinfo {author} {\bibfnamefont {K.~A.}\ \bibnamefont {Fischer}}, \bibinfo
  {author} {\bibfnamefont {S.}~\bibnamefont {Fan}},\ and\ \bibinfo {author}
  {\bibfnamefont {J.}~\bibnamefont {Vu{\v{c}}kovi{\'c}}},\ }\bibfield  {title}
  {\bibinfo {title} {Photon blockade in weakly driven cavity quantum
  electrodynamics systems with many emitters},\ }\href@noop {} {\bibfield
  {journal} {\bibinfo  {journal} {Physical review letters}\ }\textbf {\bibinfo
  {volume} {122}},\ \bibinfo {pages} {243602} (\bibinfo {year}
  {2019})}\BibitemShut {NoStop}%
\bibitem [{\citenamefont {Chakram}\ \emph {et~al.}(2022)\citenamefont
  {Chakram}, \citenamefont {He}, \citenamefont {Dixit}, \citenamefont {Oriani},
  \citenamefont {Naik}, \citenamefont {Leung}, \citenamefont {Kwon},
  \citenamefont {Ma}, \citenamefont {Jiang},\ and\ \citenamefont
  {Schuster}}]{chakram2022multimode}%
  \BibitemOpen
  \bibfield  {author} {\bibinfo {author} {\bibfnamefont {S.}~\bibnamefont
  {Chakram}}, \bibinfo {author} {\bibfnamefont {K.}~\bibnamefont {He}},
  \bibinfo {author} {\bibfnamefont {A.~V.}\ \bibnamefont {Dixit}}, \bibinfo
  {author} {\bibfnamefont {A.~E.}\ \bibnamefont {Oriani}}, \bibinfo {author}
  {\bibfnamefont {R.~K.}\ \bibnamefont {Naik}}, \bibinfo {author}
  {\bibfnamefont {N.}~\bibnamefont {Leung}}, \bibinfo {author} {\bibfnamefont
  {H.}~\bibnamefont {Kwon}}, \bibinfo {author} {\bibfnamefont {W.-L.}\
  \bibnamefont {Ma}}, \bibinfo {author} {\bibfnamefont {L.}~\bibnamefont
  {Jiang}},\ and\ \bibinfo {author} {\bibfnamefont {D.~I.}\ \bibnamefont
  {Schuster}},\ }\bibfield  {title} {\bibinfo {title} {Multimode photon
  blockade},\ }\href@noop {} {\bibfield  {journal} {\bibinfo  {journal} {Nature
  Physics}\ }\textbf {\bibinfo {volume} {18}},\ \bibinfo {pages} {879}
  (\bibinfo {year} {2022})}\BibitemShut {NoStop}%
\bibitem [{\citenamefont {Li}\ \emph {et~al.}(2022)\citenamefont {Li},
  \citenamefont {Zhang}, \citenamefont {Wu}, \citenamefont {Dong},
  \citenamefont {Zou}, \citenamefont {Guo},\ and\ \citenamefont
  {Zou}}]{li2022single}%
  \BibitemOpen
  \bibfield  {author} {\bibinfo {author} {\bibfnamefont {M.}~\bibnamefont
  {Li}}, \bibinfo {author} {\bibfnamefont {Y.-L.}\ \bibnamefont {Zhang}},
  \bibinfo {author} {\bibfnamefont {S.-H.}\ \bibnamefont {Wu}}, \bibinfo
  {author} {\bibfnamefont {C.-H.}\ \bibnamefont {Dong}}, \bibinfo {author}
  {\bibfnamefont {X.-B.}\ \bibnamefont {Zou}}, \bibinfo {author} {\bibfnamefont
  {G.-C.}\ \bibnamefont {Guo}},\ and\ \bibinfo {author} {\bibfnamefont {C.-L.}\
  \bibnamefont {Zou}},\ }\bibfield  {title} {\bibinfo {title} {Single-mode
  photon blockade enhanced by bi-tone drive},\ }\href@noop {} {\bibfield
  {journal} {\bibinfo  {journal} {Physical Review Letters}\ }\textbf {\bibinfo
  {volume} {129}},\ \bibinfo {pages} {043601} (\bibinfo {year}
  {2022})}\BibitemShut {NoStop}%
\bibitem [{\citenamefont {Jabri}\ and\ \citenamefont
  {Eleuch}(2022)}]{jabri2022enhanced}%
  \BibitemOpen
  \bibfield  {author} {\bibinfo {author} {\bibfnamefont {H.}~\bibnamefont
  {Jabri}}\ and\ \bibinfo {author} {\bibfnamefont {H.}~\bibnamefont {Eleuch}},\
  }\bibfield  {title} {\bibinfo {title} {Enhanced unconventional
  photon-blockade effect in one-and two-qubit cavities interacting with
  nonclassical light},\ }\href@noop {} {\bibfield  {journal} {\bibinfo
  {journal} {Physical Review A}\ }\textbf {\bibinfo {volume} {106}},\ \bibinfo
  {pages} {023704} (\bibinfo {year} {2022})}\BibitemShut {NoStop}%
\bibitem [{\citenamefont {Xia}\ \emph {et~al.}(2021)\citenamefont {Xia},
  \citenamefont {Zhang}, \citenamefont {Xu}, \citenamefont {Li}, \citenamefont
  {Fu},\ and\ \citenamefont {Yang}}]{xia2021giant}%
  \BibitemOpen
  \bibfield  {author} {\bibinfo {author} {\bibfnamefont {X.}~\bibnamefont
  {Xia}}, \bibinfo {author} {\bibfnamefont {X.}~\bibnamefont {Zhang}}, \bibinfo
  {author} {\bibfnamefont {J.}~\bibnamefont {Xu}}, \bibinfo {author}
  {\bibfnamefont {H.}~\bibnamefont {Li}}, \bibinfo {author} {\bibfnamefont
  {Z.}~\bibnamefont {Fu}},\ and\ \bibinfo {author} {\bibfnamefont
  {Y.}~\bibnamefont {Yang}},\ }\bibfield  {title} {\bibinfo {title} {Giant
  nonreciprocal unconventional photon blockade with a single atom in an
  asymmetric cavity},\ }\href@noop {} {\bibfield  {journal} {\bibinfo
  {journal} {Physical Review A}\ }\textbf {\bibinfo {volume} {104}},\ \bibinfo
  {pages} {063713} (\bibinfo {year} {2021})}\BibitemShut {NoStop}%
\bibitem [{\citenamefont {Wang}\ \emph {et~al.}(2021)\citenamefont {Wang},
  \citenamefont {Verstraelen}, \citenamefont {Zhang}, \citenamefont {Liew},\
  and\ \citenamefont {Chong}}]{wang2021giant}%
  \BibitemOpen
  \bibfield  {author} {\bibinfo {author} {\bibfnamefont {Y.}~\bibnamefont
  {Wang}}, \bibinfo {author} {\bibfnamefont {W.}~\bibnamefont {Verstraelen}},
  \bibinfo {author} {\bibfnamefont {B.}~\bibnamefont {Zhang}}, \bibinfo
  {author} {\bibfnamefont {T.~C.}\ \bibnamefont {Liew}},\ and\ \bibinfo
  {author} {\bibfnamefont {Y.}~\bibnamefont {Chong}},\ }\bibfield  {title}
  {\bibinfo {title} {Giant enhancement of unconventional photon blockade in a
  dimer chain},\ }\href@noop {} {\bibfield  {journal} {\bibinfo  {journal}
  {Physical Review Letters}\ }\textbf {\bibinfo {volume} {127}},\ \bibinfo
  {pages} {240402} (\bibinfo {year} {2021})}\BibitemShut {NoStop}%
\bibitem [{\citenamefont {Shen}\ \emph {et~al.}(2020)\citenamefont {Shen},
  \citenamefont {Wang}, \citenamefont {Wang},\ and\ \citenamefont
  {Yi}}]{shen2020nonreciprocal}%
  \BibitemOpen
  \bibfield  {author} {\bibinfo {author} {\bibfnamefont {H.}~\bibnamefont
  {Shen}}, \bibinfo {author} {\bibfnamefont {Q.}~\bibnamefont {Wang}}, \bibinfo
  {author} {\bibfnamefont {J.}~\bibnamefont {Wang}},\ and\ \bibinfo {author}
  {\bibfnamefont {X.}~\bibnamefont {Yi}},\ }\bibfield  {title} {\bibinfo
  {title} {Nonreciprocal unconventional photon blockade in a driven dissipative
  cavity with parametric amplification},\ }\href@noop {} {\bibfield  {journal}
  {\bibinfo  {journal} {Physical Review A}\ }\textbf {\bibinfo {volume}
  {101}},\ \bibinfo {pages} {013826} (\bibinfo {year} {2020})}\BibitemShut
  {NoStop}%
\bibitem [{\citenamefont {Hoffman}\ \emph {et~al.}(2011)\citenamefont
  {Hoffman}, \citenamefont {Srinivasan}, \citenamefont {Schmidt}, \citenamefont
  {Spietz}, \citenamefont {Aumentado}, \citenamefont {T{\"u}reci},\ and\
  \citenamefont {Houck}}]{hoffman2011dispersive}%
  \BibitemOpen
  \bibfield  {author} {\bibinfo {author} {\bibfnamefont {A.~J.}\ \bibnamefont
  {Hoffman}}, \bibinfo {author} {\bibfnamefont {S.~J.}\ \bibnamefont
  {Srinivasan}}, \bibinfo {author} {\bibfnamefont {S.}~\bibnamefont {Schmidt}},
  \bibinfo {author} {\bibfnamefont {L.}~\bibnamefont {Spietz}}, \bibinfo
  {author} {\bibfnamefont {J.}~\bibnamefont {Aumentado}}, \bibinfo {author}
  {\bibfnamefont {H.~E.}\ \bibnamefont {T{\"u}reci}},\ and\ \bibinfo {author}
  {\bibfnamefont {A.~A.}\ \bibnamefont {Houck}},\ }\bibfield  {title} {\bibinfo
  {title} {Dispersive photon blockade in a superconducting circuit},\
  }\href@noop {} {\bibfield  {journal} {\bibinfo  {journal} {Physical review
  letters}\ }\textbf {\bibinfo {volume} {107}},\ \bibinfo {pages} {053602}
  (\bibinfo {year} {2011})}\BibitemShut {NoStop}%
\bibitem [{\citenamefont {Lang}\ \emph {et~al.}(2011)\citenamefont {Lang},
  \citenamefont {Bozyigit}, \citenamefont {Eichler}, \citenamefont {Steffen},
  \citenamefont {Fink}, \citenamefont {Abdumalikov~Jr}, \citenamefont {Baur},
  \citenamefont {Filipp}, \citenamefont {Da~Silva}, \citenamefont {Blais} \emph
  {et~al.}}]{lang2011observation}%
  \BibitemOpen
  \bibfield  {author} {\bibinfo {author} {\bibfnamefont {C.}~\bibnamefont
  {Lang}}, \bibinfo {author} {\bibfnamefont {D.}~\bibnamefont {Bozyigit}},
  \bibinfo {author} {\bibfnamefont {C.}~\bibnamefont {Eichler}}, \bibinfo
  {author} {\bibfnamefont {L.}~\bibnamefont {Steffen}}, \bibinfo {author}
  {\bibfnamefont {J.}~\bibnamefont {Fink}}, \bibinfo {author} {\bibfnamefont
  {A.}~\bibnamefont {Abdumalikov~Jr}}, \bibinfo {author} {\bibfnamefont
  {M.}~\bibnamefont {Baur}}, \bibinfo {author} {\bibfnamefont {S.}~\bibnamefont
  {Filipp}}, \bibinfo {author} {\bibfnamefont {M.~P.}\ \bibnamefont
  {Da~Silva}}, \bibinfo {author} {\bibfnamefont {A.}~\bibnamefont {Blais}},
  \emph {et~al.},\ }\bibfield  {title} {\bibinfo {title} {Observation of
  resonant photon blockade at microwave frequencies using correlation function
  measurements},\ }\href@noop {} {\bibfield  {journal} {\bibinfo  {journal}
  {Physical review letters}\ }\textbf {\bibinfo {volume} {106}},\ \bibinfo
  {pages} {243601} (\bibinfo {year} {2011})}\BibitemShut {NoStop}%
\bibitem [{\citenamefont {Liu}\ \emph {et~al.}(2014)\citenamefont {Liu},
  \citenamefont {Xu}, \citenamefont {Miranowicz},\ and\ \citenamefont
  {Nori}}]{liu2014blockade}%
  \BibitemOpen
  \bibfield  {author} {\bibinfo {author} {\bibfnamefont {Y.-x.}\ \bibnamefont
  {Liu}}, \bibinfo {author} {\bibfnamefont {X.-W.}\ \bibnamefont {Xu}},
  \bibinfo {author} {\bibfnamefont {A.}~\bibnamefont {Miranowicz}},\ and\
  \bibinfo {author} {\bibfnamefont {F.}~\bibnamefont {Nori}},\ }\bibfield
  {title} {\bibinfo {title} {From blockade to transparency: Controllable photon
  transmission through a circuit-qed system},\ }\href@noop {} {\bibfield
  {journal} {\bibinfo  {journal} {Physical Review A}\ }\textbf {\bibinfo
  {volume} {89}},\ \bibinfo {pages} {043818} (\bibinfo {year}
  {2014})}\BibitemShut {NoStop}%
\bibitem [{\citenamefont {Vaneph}\ \emph {et~al.}(2018)\citenamefont {Vaneph},
  \citenamefont {Morvan}, \citenamefont {Aiello}, \citenamefont {F{\'e}chant},
  \citenamefont {Aprili}, \citenamefont {Gabelli},\ and\ \citenamefont
  {Est{\`e}ve}}]{vaneph2018observation}%
  \BibitemOpen
  \bibfield  {author} {\bibinfo {author} {\bibfnamefont {C.}~\bibnamefont
  {Vaneph}}, \bibinfo {author} {\bibfnamefont {A.}~\bibnamefont {Morvan}},
  \bibinfo {author} {\bibfnamefont {G.}~\bibnamefont {Aiello}}, \bibinfo
  {author} {\bibfnamefont {M.}~\bibnamefont {F{\'e}chant}}, \bibinfo {author}
  {\bibfnamefont {M.}~\bibnamefont {Aprili}}, \bibinfo {author} {\bibfnamefont
  {J.}~\bibnamefont {Gabelli}},\ and\ \bibinfo {author} {\bibfnamefont
  {J.}~\bibnamefont {Est{\`e}ve}},\ }\bibfield  {title} {\bibinfo {title}
  {Observation of the unconventional photon blockade in the microwave domain},\
  }\href@noop {} {\bibfield  {journal} {\bibinfo  {journal} {Physical review
  letters}\ }\textbf {\bibinfo {volume} {121}},\ \bibinfo {pages} {043602}
  (\bibinfo {year} {2018})}\BibitemShut {NoStop}%
\bibitem [{\citenamefont {Rabl}(2011)}]{rabl2011photon}%
  \BibitemOpen
  \bibfield  {author} {\bibinfo {author} {\bibfnamefont {P.}~\bibnamefont
  {Rabl}},\ }\bibfield  {title} {\bibinfo {title} {Photon blockade effect in
  optomechanical systems},\ }\href@noop {} {\bibfield  {journal} {\bibinfo
  {journal} {Physical review letters}\ }\textbf {\bibinfo {volume} {107}},\
  \bibinfo {pages} {063601} (\bibinfo {year} {2011})}\BibitemShut {NoStop}%
\bibitem [{\citenamefont {Li}\ \emph {et~al.}(2019)\citenamefont {Li},
  \citenamefont {Huang}, \citenamefont {Xu}, \citenamefont {Miranowicz},\ and\
  \citenamefont {Jing}}]{li2019nonreciprocal}%
  \BibitemOpen
  \bibfield  {author} {\bibinfo {author} {\bibfnamefont {B.}~\bibnamefont
  {Li}}, \bibinfo {author} {\bibfnamefont {R.}~\bibnamefont {Huang}}, \bibinfo
  {author} {\bibfnamefont {X.}~\bibnamefont {Xu}}, \bibinfo {author}
  {\bibfnamefont {A.}~\bibnamefont {Miranowicz}},\ and\ \bibinfo {author}
  {\bibfnamefont {H.}~\bibnamefont {Jing}},\ }\bibfield  {title} {\bibinfo
  {title} {Nonreciprocal unconventional photon blockade in a spinning
  optomechanical system},\ }\href@noop {} {\bibfield  {journal} {\bibinfo
  {journal} {Photonics Research}\ }\textbf {\bibinfo {volume} {7}},\ \bibinfo
  {pages} {630} (\bibinfo {year} {2019})}\BibitemShut {NoStop}%
\bibitem [{\citenamefont {Liao}\ \emph
  {et~al.}(2013{\natexlab{a}})\citenamefont {Liao}, \citenamefont {Nori} \emph
  {et~al.}}]{liao2013photon}%
  \BibitemOpen
  \bibfield  {author} {\bibinfo {author} {\bibfnamefont {J.-Q.}\ \bibnamefont
  {Liao}}, \bibinfo {author} {\bibfnamefont {F.}~\bibnamefont {Nori}}, \emph
  {et~al.},\ }\bibfield  {title} {\bibinfo {title} {Photon blockade in
  quadratically coupled optomechanical systems},\ }\href@noop {} {\bibfield
  {journal} {\bibinfo  {journal} {Physical Review A}\ }\textbf {\bibinfo
  {volume} {88}},\ \bibinfo {pages} {023853} (\bibinfo {year}
  {2013}{\natexlab{a}})}\BibitemShut {NoStop}%
\bibitem [{\citenamefont {Liao}\ \emph
  {et~al.}(2013{\natexlab{b}})\citenamefont {Liao}, \citenamefont {Law} \emph
  {et~al.}}]{liao2013correlated}%
  \BibitemOpen
  \bibfield  {author} {\bibinfo {author} {\bibfnamefont {J.-Q.}\ \bibnamefont
  {Liao}}, \bibinfo {author} {\bibfnamefont {C.}~\bibnamefont {Law}}, \emph
  {et~al.},\ }\bibfield  {title} {\bibinfo {title} {Correlated two-photon
  scattering in cavity optomechanics},\ }\href@noop {} {\bibfield  {journal}
  {\bibinfo  {journal} {Physical Review A}\ }\textbf {\bibinfo {volume} {87}},\
  \bibinfo {pages} {043809} (\bibinfo {year} {2013}{\natexlab{b}})}\BibitemShut
  {NoStop}%
\bibitem [{\citenamefont {Komar}\ \emph {et~al.}(2013)\citenamefont {Komar},
  \citenamefont {Bennett}, \citenamefont {Stannigel}, \citenamefont {Habraken},
  \citenamefont {Rabl}, \citenamefont {Zoller},\ and\ \citenamefont
  {Lukin}}]{komar2013single}%
  \BibitemOpen
  \bibfield  {author} {\bibinfo {author} {\bibfnamefont {P.}~\bibnamefont
  {Komar}}, \bibinfo {author} {\bibfnamefont {S.}~\bibnamefont {Bennett}},
  \bibinfo {author} {\bibfnamefont {K.}~\bibnamefont {Stannigel}}, \bibinfo
  {author} {\bibfnamefont {S.}~\bibnamefont {Habraken}}, \bibinfo {author}
  {\bibfnamefont {P.}~\bibnamefont {Rabl}}, \bibinfo {author} {\bibfnamefont
  {P.}~\bibnamefont {Zoller}},\ and\ \bibinfo {author} {\bibfnamefont {M.~D.}\
  \bibnamefont {Lukin}},\ }\bibfield  {title} {\bibinfo {title} {Single-photon
  nonlinearities in two-mode optomechanics},\ }\href@noop {} {\bibfield
  {journal} {\bibinfo  {journal} {Physical Review A}\ }\textbf {\bibinfo
  {volume} {87}},\ \bibinfo {pages} {013839} (\bibinfo {year}
  {2013})}\BibitemShut {NoStop}%
\bibitem [{\citenamefont {Sun}\ and\ \citenamefont
  {Shen}(2023)}]{sun2023photon}%
  \BibitemOpen
  \bibfield  {author} {\bibinfo {author} {\bibfnamefont {J.}~\bibnamefont
  {Sun}}\ and\ \bibinfo {author} {\bibfnamefont {H.}~\bibnamefont {Shen}},\
  }\bibfield  {title} {\bibinfo {title} {Photon blockade in non-hermitian
  optomechanical systems with nonreciprocal couplings},\ }\href@noop {}
  {\bibfield  {journal} {\bibinfo  {journal} {Physical Review A}\ }\textbf
  {\bibinfo {volume} {107}},\ \bibinfo {pages} {043715} (\bibinfo {year}
  {2023})}\BibitemShut {NoStop}%
\bibitem [{\citenamefont {Gao}\ \emph {et~al.}(2023)\citenamefont {Gao},
  \citenamefont {Cao}, \citenamefont {Lu},\ and\ \citenamefont
  {Wang}}]{gao2023phase}%
  \BibitemOpen
  \bibfield  {author} {\bibinfo {author} {\bibfnamefont {Y.-P.}\ \bibnamefont
  {Gao}}, \bibinfo {author} {\bibfnamefont {C.}~\bibnamefont {Cao}}, \bibinfo
  {author} {\bibfnamefont {P.-F.}\ \bibnamefont {Lu}},\ and\ \bibinfo {author}
  {\bibfnamefont {C.}~\bibnamefont {Wang}},\ }\bibfield  {title} {\bibinfo
  {title} {Phase-controlled photon blockade in optomechanical systems},\
  }\href@noop {} {\bibfield  {journal} {\bibinfo  {journal} {Fundamental
  Research}\ }\textbf {\bibinfo {volume} {3}},\ \bibinfo {pages} {30} (\bibinfo
  {year} {2023})}\BibitemShut {NoStop}%
\bibitem [{\citenamefont {Liu}\ \emph {et~al.}(2023)\citenamefont {Liu},
  \citenamefont {Cheng}, \citenamefont {Wang},\ and\ \citenamefont
  {Yi}}]{liu2023nonreciprocal}%
  \BibitemOpen
  \bibfield  {author} {\bibinfo {author} {\bibfnamefont {Y.-M.}\ \bibnamefont
  {Liu}}, \bibinfo {author} {\bibfnamefont {J.}~\bibnamefont {Cheng}}, \bibinfo
  {author} {\bibfnamefont {H.-F.}\ \bibnamefont {Wang}},\ and\ \bibinfo
  {author} {\bibfnamefont {X.}~\bibnamefont {Yi}},\ }\bibfield  {title}
  {\bibinfo {title} {Nonreciprocal photon blockade in a spinning optomechanical
  system with nonreciprocal coupling},\ }\href@noop {} {\bibfield  {journal}
  {\bibinfo  {journal} {Optics Express}\ }\textbf {\bibinfo {volume} {31}},\
  \bibinfo {pages} {12847} (\bibinfo {year} {2023})}\BibitemShut {NoStop}%
\bibitem [{\citenamefont {Li}\ \emph {et~al.}(2024)\citenamefont {Li},
  \citenamefont {Hu},\ and\ \citenamefont {Yang}}]{li2024enhancement}%
  \BibitemOpen
  \bibfield  {author} {\bibinfo {author} {\bibfnamefont {J.}~\bibnamefont
  {Li}}, \bibinfo {author} {\bibfnamefont {C.-M.}\ \bibnamefont {Hu}},\ and\
  \bibinfo {author} {\bibfnamefont {Y.}~\bibnamefont {Yang}},\ }\bibfield
  {title} {\bibinfo {title} {Enhancement of photon blockade via topological
  edge states},\ }\href@noop {} {\bibfield  {journal} {\bibinfo  {journal}
  {Physical Review Applied}\ }\textbf {\bibinfo {volume} {21}},\ \bibinfo
  {pages} {034058} (\bibinfo {year} {2024})}\BibitemShut {NoStop}%
\bibitem [{\citenamefont {Zhang}\ \emph {et~al.}(2023)\citenamefont {Zhang},
  \citenamefont {Wang}, \citenamefont {Liu}, \citenamefont {Zhang},\ and\
  \citenamefont {Wang}}]{zhang2023nonreciprocal}%
  \BibitemOpen
  \bibfield  {author} {\bibinfo {author} {\bibfnamefont {W.}~\bibnamefont
  {Zhang}}, \bibinfo {author} {\bibfnamefont {T.}~\bibnamefont {Wang}},
  \bibinfo {author} {\bibfnamefont {S.}~\bibnamefont {Liu}}, \bibinfo {author}
  {\bibfnamefont {S.}~\bibnamefont {Zhang}},\ and\ \bibinfo {author}
  {\bibfnamefont {H.-F.}\ \bibnamefont {Wang}},\ }\bibfield  {title} {\bibinfo
  {title} {Nonreciprocal photon blockade in a spinning resonator coupled to two
  two-level atoms},\ }\href@noop {} {\bibfield  {journal} {\bibinfo  {journal}
  {Science China Physics, Mechanics \& Astronomy}\ }\textbf {\bibinfo {volume}
  {66}},\ \bibinfo {pages} {240313} (\bibinfo {year} {2023})}\BibitemShut
  {NoStop}%
\bibitem [{\citenamefont {Lu}\ \emph {et~al.}(2025)\citenamefont {Lu},
  \citenamefont {Wu},\ and\ \citenamefont {L{\"u}}}]{lu2025chiral}%
  \BibitemOpen
  \bibfield  {author} {\bibinfo {author} {\bibfnamefont {Z.-G.}\ \bibnamefont
  {Lu}}, \bibinfo {author} {\bibfnamefont {Y.}~\bibnamefont {Wu}},\ and\
  \bibinfo {author} {\bibfnamefont {X.-Y.}\ \bibnamefont {L{\"u}}},\ }\bibfield
   {title} {\bibinfo {title} {Chiral interaction induced near-perfect photon
  blockade},\ }\href@noop {} {\bibfield  {journal} {\bibinfo  {journal}
  {Physical Review Letters}\ }\textbf {\bibinfo {volume} {134}},\ \bibinfo
  {pages} {013602} (\bibinfo {year} {2025})}\BibitemShut {NoStop}%
\bibitem [{\citenamefont {Xie}\ \emph {et~al.}(2022)\citenamefont {Xie},
  \citenamefont {He}, \citenamefont {Shang}, \citenamefont {Lin},\ and\
  \citenamefont {Lin}}]{xie2022nonreciprocal}%
  \BibitemOpen
  \bibfield  {author} {\bibinfo {author} {\bibfnamefont {H.}~\bibnamefont
  {Xie}}, \bibinfo {author} {\bibfnamefont {L.-W.}\ \bibnamefont {He}},
  \bibinfo {author} {\bibfnamefont {X.}~\bibnamefont {Shang}}, \bibinfo
  {author} {\bibfnamefont {G.-W.}\ \bibnamefont {Lin}},\ and\ \bibinfo {author}
  {\bibfnamefont {X.-M.}\ \bibnamefont {Lin}},\ }\bibfield  {title} {\bibinfo
  {title} {Nonreciprocal photon blockade in cavity optomagnonics},\ }\href@noop
  {} {\bibfield  {journal} {\bibinfo  {journal} {Physical Review A}\ }\textbf
  {\bibinfo {volume} {106}},\ \bibinfo {pages} {053707} (\bibinfo {year}
  {2022})}\BibitemShut {NoStop}%
\bibitem [{\citenamefont {Fan}\ \emph {et~al.}(2024)\citenamefont {Fan},
  \citenamefont {Zhang}, \citenamefont {Yu}, \citenamefont {Liu}, \citenamefont
  {He}, \citenamefont {Li},\ and\ \citenamefont
  {Xiong}}]{fan2024nonreciprocal}%
  \BibitemOpen
  \bibfield  {author} {\bibinfo {author} {\bibfnamefont {X.-H.}\ \bibnamefont
  {Fan}}, \bibinfo {author} {\bibfnamefont {Y.-N.}\ \bibnamefont {Zhang}},
  \bibinfo {author} {\bibfnamefont {J.-P.}\ \bibnamefont {Yu}}, \bibinfo
  {author} {\bibfnamefont {M.-Y.}\ \bibnamefont {Liu}}, \bibinfo {author}
  {\bibfnamefont {W.-D.}\ \bibnamefont {He}}, \bibinfo {author} {\bibfnamefont
  {H.-C.}\ \bibnamefont {Li}},\ and\ \bibinfo {author} {\bibfnamefont
  {W.}~\bibnamefont {Xiong}},\ }\bibfield  {title} {\bibinfo {title}
  {Nonreciprocal unconventional photon blockade with kerr magnons},\
  }\href@noop {} {\bibfield  {journal} {\bibinfo  {journal} {Advanced Quantum
  Technologies}\ }\textbf {\bibinfo {volume} {7}},\ \bibinfo {pages} {2400043}
  (\bibinfo {year} {2024})}\BibitemShut {NoStop}%
\bibitem [{\citenamefont {Miranowicz}\ \emph {et~al.}(2013)\citenamefont
  {Miranowicz}, \citenamefont {Paprzycka}, \citenamefont {Liu}, \citenamefont
  {Bajer},\ and\ \citenamefont {Nori}}]{miranowicz2013two}%
  \BibitemOpen
  \bibfield  {author} {\bibinfo {author} {\bibfnamefont {A.}~\bibnamefont
  {Miranowicz}}, \bibinfo {author} {\bibfnamefont {M.}~\bibnamefont
  {Paprzycka}}, \bibinfo {author} {\bibfnamefont {Y.-x.}\ \bibnamefont {Liu}},
  \bibinfo {author} {\bibfnamefont {J.}~\bibnamefont {Bajer}},\ and\ \bibinfo
  {author} {\bibfnamefont {F.}~\bibnamefont {Nori}},\ }\bibfield  {title}
  {\bibinfo {title} {Two-photon and three-photon blockades in driven nonlinear
  systems},\ }\href@noop {} {\bibfield  {journal} {\bibinfo  {journal}
  {Physical Review A}\ }\textbf {\bibinfo {volume} {87}},\ \bibinfo {pages}
  {023809} (\bibinfo {year} {2013})}\BibitemShut {NoStop}%
\bibitem [{\citenamefont {Hovsepyan}\ \emph {et~al.}(2014)\citenamefont
  {Hovsepyan}, \citenamefont {Shahinyan},\ and\ \citenamefont
  {Kryuchkyan}}]{hovsepyan2014multiphoton}%
  \BibitemOpen
  \bibfield  {author} {\bibinfo {author} {\bibfnamefont {G.}~\bibnamefont
  {Hovsepyan}}, \bibinfo {author} {\bibfnamefont {A.}~\bibnamefont
  {Shahinyan}},\ and\ \bibinfo {author} {\bibfnamefont {G.~Y.}\ \bibnamefont
  {Kryuchkyan}},\ }\bibfield  {title} {\bibinfo {title} {Multiphoton blockades
  in pulsed regimes beyond stationary limits},\ }\href@noop {} {\bibfield
  {journal} {\bibinfo  {journal} {Physical Review A}\ }\textbf {\bibinfo
  {volume} {90}},\ \bibinfo {pages} {013839} (\bibinfo {year}
  {2014})}\BibitemShut {NoStop}%
\bibitem [{\citenamefont {Deng}\ \emph {et~al.}(2015)\citenamefont {Deng},
  \citenamefont {Li},\ and\ \citenamefont {Qin}}]{deng2015enhancement}%
  \BibitemOpen
  \bibfield  {author} {\bibinfo {author} {\bibfnamefont {W.-W.}\ \bibnamefont
  {Deng}}, \bibinfo {author} {\bibfnamefont {G.-X.}\ \bibnamefont {Li}},\ and\
  \bibinfo {author} {\bibfnamefont {H.}~\bibnamefont {Qin}},\ }\bibfield
  {title} {\bibinfo {title} {Enhancement of the two-photon blockade in a
  strong-coupling qubit-cavity system},\ }\href@noop {} {\bibfield  {journal}
  {\bibinfo  {journal} {Physical Review A}\ }\textbf {\bibinfo {volume} {91}},\
  \bibinfo {pages} {043831} (\bibinfo {year} {2015})}\BibitemShut {NoStop}%
\bibitem [{\citenamefont {Zhu}\ \emph {et~al.}(2017)\citenamefont {Zhu},
  \citenamefont {Yang},\ and\ \citenamefont {Agarwal}}]{zhu2017collective}%
  \BibitemOpen
  \bibfield  {author} {\bibinfo {author} {\bibfnamefont {C.}~\bibnamefont
  {Zhu}}, \bibinfo {author} {\bibfnamefont {Y.}~\bibnamefont {Yang}},\ and\
  \bibinfo {author} {\bibfnamefont {G.}~\bibnamefont {Agarwal}},\ }\bibfield
  {title} {\bibinfo {title} {Collective multiphoton blockade in cavity quantum
  electrodynamics},\ }\href@noop {} {\bibfield  {journal} {\bibinfo  {journal}
  {Physical Review A}\ }\textbf {\bibinfo {volume} {95}},\ \bibinfo {pages}
  {063842} (\bibinfo {year} {2017})}\BibitemShut {NoStop}%
\bibitem [{\citenamefont {Bin}\ \emph {et~al.}(2018)\citenamefont {Bin},
  \citenamefont {L{\"u}}, \citenamefont {Bin},\ and\ \citenamefont
  {Wu}}]{bin2018two}%
  \BibitemOpen
  \bibfield  {author} {\bibinfo {author} {\bibfnamefont {Q.}~\bibnamefont
  {Bin}}, \bibinfo {author} {\bibfnamefont {X.-Y.}\ \bibnamefont {L{\"u}}},
  \bibinfo {author} {\bibfnamefont {S.-W.}\ \bibnamefont {Bin}},\ and\ \bibinfo
  {author} {\bibfnamefont {Y.}~\bibnamefont {Wu}},\ }\bibfield  {title}
  {\bibinfo {title} {Two-photon blockade in a cascaded
  cavity-quantum-electrodynamics system},\ }\href@noop {} {\bibfield  {journal}
  {\bibinfo  {journal} {Physical Review A}\ }\textbf {\bibinfo {volume} {98}},\
  \bibinfo {pages} {043858} (\bibinfo {year} {2018})}\BibitemShut {NoStop}%
\bibitem [{\citenamefont {Kowalewska-Kud{\l}aszyk}\ \emph
  {et~al.}(2019)\citenamefont {Kowalewska-Kud{\l}aszyk}, \citenamefont {Abo},
  \citenamefont {Chimczak}, \citenamefont {Pe{\v{r}}ina~Jr}, \citenamefont
  {Nori},\ and\ \citenamefont {Miranowicz}}]{kowalewska2019two}%
  \BibitemOpen
  \bibfield  {author} {\bibinfo {author} {\bibfnamefont {A.}~\bibnamefont
  {Kowalewska-Kud{\l}aszyk}}, \bibinfo {author} {\bibfnamefont {S.~I.}\
  \bibnamefont {Abo}}, \bibinfo {author} {\bibfnamefont {G.}~\bibnamefont
  {Chimczak}}, \bibinfo {author} {\bibfnamefont {J.}~\bibnamefont
  {Pe{\v{r}}ina~Jr}}, \bibinfo {author} {\bibfnamefont {F.}~\bibnamefont
  {Nori}},\ and\ \bibinfo {author} {\bibfnamefont {A.}~\bibnamefont
  {Miranowicz}},\ }\bibfield  {title} {\bibinfo {title} {Two-photon blockade
  and photon-induced tunneling generated by squeezing},\ }\href@noop {}
  {\bibfield  {journal} {\bibinfo  {journal} {Physical Review A}\ }\textbf
  {\bibinfo {volume} {100}},\ \bibinfo {pages} {053857} (\bibinfo {year}
  {2019})}\BibitemShut {NoStop}%
\bibitem [{\citenamefont {Hamsen}\ \emph {et~al.}(2017)\citenamefont {Hamsen},
  \citenamefont {Tolazzi}, \citenamefont {Wilk},\ and\ \citenamefont
  {Rempe}}]{hamsen2017two}%
  \BibitemOpen
  \bibfield  {author} {\bibinfo {author} {\bibfnamefont {C.}~\bibnamefont
  {Hamsen}}, \bibinfo {author} {\bibfnamefont {K.~N.}\ \bibnamefont {Tolazzi}},
  \bibinfo {author} {\bibfnamefont {T.}~\bibnamefont {Wilk}},\ and\ \bibinfo
  {author} {\bibfnamefont {G.}~\bibnamefont {Rempe}},\ }\bibfield  {title}
  {\bibinfo {title} {Two-photon blockade in an atom-driven cavity qed system},\
  }\href@noop {} {\bibfield  {journal} {\bibinfo  {journal} {Physical review
  letters}\ }\textbf {\bibinfo {volume} {118}},\ \bibinfo {pages} {133604}
  (\bibinfo {year} {2017})}\BibitemShut {NoStop}%
\bibitem [{\citenamefont {Zhou}\ \emph {et~al.}(2021)\citenamefont {Zhou},
  \citenamefont {Minganti}, \citenamefont {Qin}, \citenamefont {Wu},
  \citenamefont {Zhao}, \citenamefont {Fang}, \citenamefont {Nori},\ and\
  \citenamefont {Yang}}]{zhou2021n}%
  \BibitemOpen
  \bibfield  {author} {\bibinfo {author} {\bibfnamefont {Y.~H.}\ \bibnamefont
  {Zhou}}, \bibinfo {author} {\bibfnamefont {F.}~\bibnamefont {Minganti}},
  \bibinfo {author} {\bibfnamefont {W.}~\bibnamefont {Qin}}, \bibinfo {author}
  {\bibfnamefont {Q.-C.}\ \bibnamefont {Wu}}, \bibinfo {author} {\bibfnamefont
  {J.-L.}\ \bibnamefont {Zhao}}, \bibinfo {author} {\bibfnamefont {Y.-L.}\
  \bibnamefont {Fang}}, \bibinfo {author} {\bibfnamefont {F.}~\bibnamefont
  {Nori}},\ and\ \bibinfo {author} {\bibfnamefont {C.-P.}\ \bibnamefont
  {Yang}},\ }\bibfield  {title} {\bibinfo {title} {n-photon blockade with an
  n-photon parametric drive},\ }\href@noop {} {\bibfield  {journal} {\bibinfo
  {journal} {Physical Review A}\ }\textbf {\bibinfo {volume} {104}},\ \bibinfo
  {pages} {053718} (\bibinfo {year} {2021})}\BibitemShut {NoStop}%
\bibitem [{\citenamefont {Singh}\ \emph {et~al.}(2019)\citenamefont {Singh},
  \citenamefont {Jha},\ and\ \citenamefont {Bhattacherjee}}]{singh2019photon}%
  \BibitemOpen
  \bibfield  {author} {\bibinfo {author} {\bibfnamefont {M.~K.}\ \bibnamefont
  {Singh}}, \bibinfo {author} {\bibfnamefont {P.~K.}\ \bibnamefont {Jha}},\
  and\ \bibinfo {author} {\bibfnamefont {A.~B.}\ \bibnamefont
  {Bhattacherjee}},\ }\bibfield  {title} {\bibinfo {title} {Photon blockade
  induced tunable source of one/two photon in a double quantum
  dot-semiconductor microcavity system},\ }\href@noop {} {\bibfield  {journal}
  {\bibinfo  {journal} {Optik}\ }\textbf {\bibinfo {volume} {185}},\ \bibinfo
  {pages} {685} (\bibinfo {year} {2019})}\BibitemShut {NoStop}%
\bibitem [{\citenamefont {Qu}\ \emph {et~al.}(2020)\citenamefont {Qu},
  \citenamefont {Shen}, \citenamefont {Li},\ and\ \citenamefont
  {Wu}}]{qu2020improving}%
  \BibitemOpen
  \bibfield  {author} {\bibinfo {author} {\bibfnamefont {Y.}~\bibnamefont
  {Qu}}, \bibinfo {author} {\bibfnamefont {S.}~\bibnamefont {Shen}}, \bibinfo
  {author} {\bibfnamefont {J.}~\bibnamefont {Li}},\ and\ \bibinfo {author}
  {\bibfnamefont {Y.}~\bibnamefont {Wu}},\ }\bibfield  {title} {\bibinfo
  {title} {Improving photon antibunching with two dipole-coupled atoms in
  whispering-gallery-mode microresonators},\ }\href@noop {} {\bibfield
  {journal} {\bibinfo  {journal} {Physical Review A}\ }\textbf {\bibinfo
  {volume} {101}},\ \bibinfo {pages} {023810} (\bibinfo {year}
  {2020})}\BibitemShut {NoStop}%
\bibitem [{\citenamefont {Devi}\ \emph {et~al.}(2020)\citenamefont {Devi},
  \citenamefont {Gunapala}, \citenamefont {Stockman},\ and\ \citenamefont
  {Premaratne}}]{devi2020nonequilibrium}%
  \BibitemOpen
  \bibfield  {author} {\bibinfo {author} {\bibfnamefont {A.}~\bibnamefont
  {Devi}}, \bibinfo {author} {\bibfnamefont {S.~D.}\ \bibnamefont {Gunapala}},
  \bibinfo {author} {\bibfnamefont {M.~I.}\ \bibnamefont {Stockman}},\ and\
  \bibinfo {author} {\bibfnamefont {M.}~\bibnamefont {Premaratne}},\ }\bibfield
   {title} {\bibinfo {title} {Nonequilibrium cavity qed model accounting for
  dipole-dipole interaction in strong-, ultrastrong-, and deep-strong-coupling
  regimes},\ }\href@noop {} {\bibfield  {journal} {\bibinfo  {journal}
  {Physical Review A}\ }\textbf {\bibinfo {volume} {102}},\ \bibinfo {pages}
  {013701} (\bibinfo {year} {2020})}\BibitemShut {NoStop}%
\bibitem [{\citenamefont {Williamson}\ \emph {et~al.}(2020)\citenamefont
  {Williamson}, \citenamefont {Borgh},\ and\ \citenamefont
  {Ruostekoski}}]{williamson2020superatom}%
  \BibitemOpen
  \bibfield  {author} {\bibinfo {author} {\bibfnamefont {L.}~\bibnamefont
  {Williamson}}, \bibinfo {author} {\bibfnamefont {M.~O.}\ \bibnamefont
  {Borgh}},\ and\ \bibinfo {author} {\bibfnamefont {J.}~\bibnamefont
  {Ruostekoski}},\ }\bibfield  {title} {\bibinfo {title} {Superatom picture of
  collective nonclassical light emission and dipole blockade in atom arrays},\
  }\href@noop {} {\bibfield  {journal} {\bibinfo  {journal} {Physical review
  letters}\ }\textbf {\bibinfo {volume} {125}},\ \bibinfo {pages} {073602}
  (\bibinfo {year} {2020})}\BibitemShut {NoStop}%
\bibitem [{\citenamefont {Cidrim}\ \emph {et~al.}(2020)\citenamefont {Cidrim},
  \citenamefont {do~Espirito~Santo}, \citenamefont {Schachenmayer},
  \citenamefont {Kaiser},\ and\ \citenamefont {Bachelard}}]{cidrim2020photon}%
  \BibitemOpen
  \bibfield  {author} {\bibinfo {author} {\bibfnamefont {A.}~\bibnamefont
  {Cidrim}}, \bibinfo {author} {\bibfnamefont {T.}~\bibnamefont
  {do~Espirito~Santo}}, \bibinfo {author} {\bibfnamefont {J.}~\bibnamefont
  {Schachenmayer}}, \bibinfo {author} {\bibfnamefont {R.}~\bibnamefont
  {Kaiser}},\ and\ \bibinfo {author} {\bibfnamefont {R.}~\bibnamefont
  {Bachelard}},\ }\bibfield  {title} {\bibinfo {title} {Photon blockade with
  ground-state neutral atoms},\ }\href@noop {} {\bibfield  {journal} {\bibinfo
  {journal} {Physical Review Letters}\ }\textbf {\bibinfo {volume} {125}},\
  \bibinfo {pages} {073601} (\bibinfo {year} {2020})}\BibitemShut {NoStop}%
\bibitem [{\citenamefont {Zhu}\ \emph {et~al.}(2021)\citenamefont {Zhu},
  \citenamefont {Hou}, \citenamefont {Yang},\ and\ \citenamefont
  {Deng}}]{zhu2021hybrid}%
  \BibitemOpen
  \bibfield  {author} {\bibinfo {author} {\bibfnamefont {C.}~\bibnamefont
  {Zhu}}, \bibinfo {author} {\bibfnamefont {K.}~\bibnamefont {Hou}}, \bibinfo
  {author} {\bibfnamefont {Y.}~\bibnamefont {Yang}},\ and\ \bibinfo {author}
  {\bibfnamefont {L.}~\bibnamefont {Deng}},\ }\bibfield  {title} {\bibinfo
  {title} {Hybrid level anharmonicity and interference-induced photon blockade
  in a two-qubit cavity qed system with dipole--dipole interaction},\
  }\href@noop {} {\bibfield  {journal} {\bibinfo  {journal} {Photonics
  Research}\ }\textbf {\bibinfo {volume} {9}},\ \bibinfo {pages} {1264}
  (\bibinfo {year} {2021})}\BibitemShut {NoStop}%
\bibitem [{\citenamefont {Zheng}\ \emph {et~al.}(2016)\citenamefont {Zheng},
  \citenamefont {Hu}, \citenamefont {Yang},\ and\ \citenamefont
  {Wu}}]{zheng2016photon}%
  \BibitemOpen
  \bibfield  {author} {\bibinfo {author} {\bibfnamefont {Y.-M.}\ \bibnamefont
  {Zheng}}, \bibinfo {author} {\bibfnamefont {C.-S.}\ \bibnamefont {Hu}},
  \bibinfo {author} {\bibfnamefont {Z.-B.}\ \bibnamefont {Yang}},\ and\
  \bibinfo {author} {\bibfnamefont {H.-Z.}\ \bibnamefont {Wu}},\ }\bibfield
  {title} {\bibinfo {title} {Photon bunching and anti-bunching with two
  dipole-coupled atoms in an optical cavity},\ }\href@noop {} {\bibfield
  {journal} {\bibinfo  {journal} {Chinese Physics B}\ }\textbf {\bibinfo
  {volume} {25}},\ \bibinfo {pages} {104202} (\bibinfo {year}
  {2016})}\BibitemShut {NoStop}%
\bibitem [{\citenamefont {Lukin}\ \emph {et~al.}(2001)\citenamefont {Lukin},
  \citenamefont {Fleischhauer}, \citenamefont {Cote}, \citenamefont {Duan},
  \citenamefont {Jaksch}, \citenamefont {Cirac},\ and\ \citenamefont
  {Zoller}}]{lukin2001dipole}%
  \BibitemOpen
  \bibfield  {author} {\bibinfo {author} {\bibfnamefont {M.~D.}\ \bibnamefont
  {Lukin}}, \bibinfo {author} {\bibfnamefont {M.}~\bibnamefont {Fleischhauer}},
  \bibinfo {author} {\bibfnamefont {R.}~\bibnamefont {Cote}}, \bibinfo {author}
  {\bibfnamefont {L.}~\bibnamefont {Duan}}, \bibinfo {author} {\bibfnamefont
  {D.}~\bibnamefont {Jaksch}}, \bibinfo {author} {\bibfnamefont {J.~I.}\
  \bibnamefont {Cirac}},\ and\ \bibinfo {author} {\bibfnamefont
  {P.}~\bibnamefont {Zoller}},\ }\bibfield  {title} {\bibinfo {title} {Dipole
  blockade and quantum information processing in mesoscopic atomic ensembles},\
  }\href@noop {} {\bibfield  {journal} {\bibinfo  {journal} {Physical review
  letters}\ }\textbf {\bibinfo {volume} {87}},\ \bibinfo {pages} {037901}
  (\bibinfo {year} {2001})}\BibitemShut {NoStop}%
\bibitem [{\citenamefont {Urban}\ \emph {et~al.}(2009)\citenamefont {Urban},
  \citenamefont {Johnson}, \citenamefont {Henage}, \citenamefont {Isenhower},
  \citenamefont {Yavuz}, \citenamefont {Walker},\ and\ \citenamefont
  {Saffman}}]{urban2009observation}%
  \BibitemOpen
  \bibfield  {author} {\bibinfo {author} {\bibfnamefont {E.}~\bibnamefont
  {Urban}}, \bibinfo {author} {\bibfnamefont {T.~A.}\ \bibnamefont {Johnson}},
  \bibinfo {author} {\bibfnamefont {T.}~\bibnamefont {Henage}}, \bibinfo
  {author} {\bibfnamefont {L.}~\bibnamefont {Isenhower}}, \bibinfo {author}
  {\bibfnamefont {D.}~\bibnamefont {Yavuz}}, \bibinfo {author} {\bibfnamefont
  {T.}~\bibnamefont {Walker}},\ and\ \bibinfo {author} {\bibfnamefont
  {M.}~\bibnamefont {Saffman}},\ }\bibfield  {title} {\bibinfo {title}
  {Observation of rydberg blockade between two atoms},\ }\href@noop {}
  {\bibfield  {journal} {\bibinfo  {journal} {Nature Physics}\ }\textbf
  {\bibinfo {volume} {5}},\ \bibinfo {pages} {110} (\bibinfo {year}
  {2009})}\BibitemShut {NoStop}%
\bibitem [{\citenamefont {Ga{\"e}tan}\ \emph {et~al.}(2009)\citenamefont
  {Ga{\"e}tan}, \citenamefont {Miroshnychenko}, \citenamefont {Wilk},
  \citenamefont {Chotia}, \citenamefont {Viteau}, \citenamefont {Comparat},
  \citenamefont {Pillet}, \citenamefont {Browaeys},\ and\ \citenamefont
  {Grangier}}]{gaetan2009observation}%
  \BibitemOpen
  \bibfield  {author} {\bibinfo {author} {\bibfnamefont {A.}~\bibnamefont
  {Ga{\"e}tan}}, \bibinfo {author} {\bibfnamefont {Y.}~\bibnamefont
  {Miroshnychenko}}, \bibinfo {author} {\bibfnamefont {T.}~\bibnamefont
  {Wilk}}, \bibinfo {author} {\bibfnamefont {A.}~\bibnamefont {Chotia}},
  \bibinfo {author} {\bibfnamefont {M.}~\bibnamefont {Viteau}}, \bibinfo
  {author} {\bibfnamefont {D.}~\bibnamefont {Comparat}}, \bibinfo {author}
  {\bibfnamefont {P.}~\bibnamefont {Pillet}}, \bibinfo {author} {\bibfnamefont
  {A.}~\bibnamefont {Browaeys}},\ and\ \bibinfo {author} {\bibfnamefont
  {P.}~\bibnamefont {Grangier}},\ }\bibfield  {title} {\bibinfo {title}
  {Observation of collective excitation of two individual atoms in the rydberg
  blockade regime},\ }\href@noop {} {\bibfield  {journal} {\bibinfo  {journal}
  {Nature Physics}\ }\textbf {\bibinfo {volume} {5}},\ \bibinfo {pages} {115}
  (\bibinfo {year} {2009})}\BibitemShut {NoStop}%
\bibitem [{\citenamefont {Ates}\ \emph {et~al.}(2007)\citenamefont {Ates},
  \citenamefont {Pohl}, \citenamefont {Pattard},\ and\ \citenamefont
  {Rost}}]{ates2007antiblockade}%
  \BibitemOpen
  \bibfield  {author} {\bibinfo {author} {\bibfnamefont {C.}~\bibnamefont
  {Ates}}, \bibinfo {author} {\bibfnamefont {T.}~\bibnamefont {Pohl}}, \bibinfo
  {author} {\bibfnamefont {T.}~\bibnamefont {Pattard}},\ and\ \bibinfo {author}
  {\bibfnamefont {J.~M.}\ \bibnamefont {Rost}},\ }\bibfield  {title} {\bibinfo
  {title} {Antiblockade in rydberg excitation of an ultracold lattice gas},\
  }\href@noop {} {\bibfield  {journal} {\bibinfo  {journal} {Physical review
  letters}\ }\textbf {\bibinfo {volume} {98}},\ \bibinfo {pages} {023002}
  (\bibinfo {year} {2007})}\BibitemShut {NoStop}%
\bibitem [{\citenamefont {Amthor}\ \emph {et~al.}(2010)\citenamefont {Amthor},
  \citenamefont {Giese}, \citenamefont {Hofmann},\ and\ \citenamefont
  {Weidem{\"u}ller}}]{amthor2010evidence}%
  \BibitemOpen
  \bibfield  {author} {\bibinfo {author} {\bibfnamefont {T.}~\bibnamefont
  {Amthor}}, \bibinfo {author} {\bibfnamefont {C.}~\bibnamefont {Giese}},
  \bibinfo {author} {\bibfnamefont {C.~S.}\ \bibnamefont {Hofmann}},\ and\
  \bibinfo {author} {\bibfnamefont {M.}~\bibnamefont {Weidem{\"u}ller}},\
  }\bibfield  {title} {\bibinfo {title} {Evidence of antiblockade in an
  ultracold rydberg gas},\ }\href@noop {} {\bibfield  {journal} {\bibinfo
  {journal} {Physical review letters}\ }\textbf {\bibinfo {volume} {104}},\
  \bibinfo {pages} {013001} (\bibinfo {year} {2010})}\BibitemShut {NoStop}%
\bibitem [{\citenamefont {Yan}\ \emph {et~al.}(2023)\citenamefont {Yan},
  \citenamefont {Ho}, \citenamefont {Lu}, \citenamefont {Masson}, \citenamefont
  {Asenjo-Garcia},\ and\ \citenamefont {Stamper-Kurn}}]{yan2023superradiant}%
  \BibitemOpen
  \bibfield  {author} {\bibinfo {author} {\bibfnamefont {Z.}~\bibnamefont
  {Yan}}, \bibinfo {author} {\bibfnamefont {J.}~\bibnamefont {Ho}}, \bibinfo
  {author} {\bibfnamefont {Y.-H.}\ \bibnamefont {Lu}}, \bibinfo {author}
  {\bibfnamefont {S.~J.}\ \bibnamefont {Masson}}, \bibinfo {author}
  {\bibfnamefont {A.}~\bibnamefont {Asenjo-Garcia}},\ and\ \bibinfo {author}
  {\bibfnamefont {D.~M.}\ \bibnamefont {Stamper-Kurn}},\ }\bibfield  {title}
  {\bibinfo {title} {Superradiant and subradiant cavity scattering by atom
  arrays},\ }\href@noop {} {\bibfield  {journal} {\bibinfo  {journal} {Physical
  Review Letters}\ }\textbf {\bibinfo {volume} {131}},\ \bibinfo {pages}
  {253603} (\bibinfo {year} {2023})}\BibitemShut {NoStop}%
\bibitem [{\citenamefont {Ho}\ \emph {et~al.}(2025)\citenamefont {Ho},
  \citenamefont {Lu}, \citenamefont {Xiang}, \citenamefont {Rusconi},
  \citenamefont {Masson}, \citenamefont {Asenjo-Garcia}, \citenamefont {Yan},\
  and\ \citenamefont {Stamper-Kurn}}]{ho2025optomechanical}%
  \BibitemOpen
  \bibfield  {author} {\bibinfo {author} {\bibfnamefont {J.}~\bibnamefont
  {Ho}}, \bibinfo {author} {\bibfnamefont {Y.-H.}\ \bibnamefont {Lu}}, \bibinfo
  {author} {\bibfnamefont {T.}~\bibnamefont {Xiang}}, \bibinfo {author}
  {\bibfnamefont {C.~C.}\ \bibnamefont {Rusconi}}, \bibinfo {author}
  {\bibfnamefont {S.~J.}\ \bibnamefont {Masson}}, \bibinfo {author}
  {\bibfnamefont {A.}~\bibnamefont {Asenjo-Garcia}}, \bibinfo {author}
  {\bibfnamefont {Z.}~\bibnamefont {Yan}},\ and\ \bibinfo {author}
  {\bibfnamefont {D.~M.}\ \bibnamefont {Stamper-Kurn}},\ }\bibfield  {title}
  {\bibinfo {title} {Optomechanical self-organization in a mesoscopic atom
  array},\ }\href@noop {} {\bibfield  {journal} {\bibinfo  {journal} {Nature
  Physics}\ ,\ \bibinfo {pages} {1}} (\bibinfo {year} {2025})}\BibitemShut
  {NoStop}%
\bibitem [{\citenamefont {Browaeys}\ and\ \citenamefont
  {Lahaye}(2020)}]{browaeys2020many}%
  \BibitemOpen
  \bibfield  {author} {\bibinfo {author} {\bibfnamefont {A.}~\bibnamefont
  {Browaeys}}\ and\ \bibinfo {author} {\bibfnamefont {T.}~\bibnamefont
  {Lahaye}},\ }\bibfield  {title} {\bibinfo {title} {Many-body physics with
  individually controlled rydberg atoms},\ }\href@noop {} {\bibfield  {journal}
  {\bibinfo  {journal} {Nature Physics}\ }\textbf {\bibinfo {volume} {16}},\
  \bibinfo {pages} {132} (\bibinfo {year} {2020})}\BibitemShut {NoStop}%
\bibitem [{\citenamefont {Bluvstein}\ \emph {et~al.}(2024)\citenamefont
  {Bluvstein}, \citenamefont {Evered}, \citenamefont {Geim}, \citenamefont
  {Li}, \citenamefont {Zhou}, \citenamefont {Manovitz}, \citenamefont {Ebadi},
  \citenamefont {Cain}, \citenamefont {Kalinowski}, \citenamefont {Hangleiter}
  \emph {et~al.}}]{bluvstein2024logical}%
  \BibitemOpen
  \bibfield  {author} {\bibinfo {author} {\bibfnamefont {D.}~\bibnamefont
  {Bluvstein}}, \bibinfo {author} {\bibfnamefont {S.~J.}\ \bibnamefont
  {Evered}}, \bibinfo {author} {\bibfnamefont {A.~A.}\ \bibnamefont {Geim}},
  \bibinfo {author} {\bibfnamefont {S.~H.}\ \bibnamefont {Li}}, \bibinfo
  {author} {\bibfnamefont {H.}~\bibnamefont {Zhou}}, \bibinfo {author}
  {\bibfnamefont {T.}~\bibnamefont {Manovitz}}, \bibinfo {author}
  {\bibfnamefont {S.}~\bibnamefont {Ebadi}}, \bibinfo {author} {\bibfnamefont
  {M.}~\bibnamefont {Cain}}, \bibinfo {author} {\bibfnamefont {M.}~\bibnamefont
  {Kalinowski}}, \bibinfo {author} {\bibfnamefont {D.}~\bibnamefont
  {Hangleiter}}, \emph {et~al.},\ }\bibfield  {title} {\bibinfo {title}
  {Logical quantum processor based on reconfigurable atom arrays},\ }\href@noop
  {} {\bibfield  {journal} {\bibinfo  {journal} {Nature}\ }\textbf {\bibinfo
  {volume} {626}},\ \bibinfo {pages} {58} (\bibinfo {year} {2024})}\BibitemShut
  {NoStop}%
\bibitem [{\citenamefont {Morgan}\ and\ \citenamefont
  {Hogan}(2020)}]{morgan2020coupling}%
  \BibitemOpen
  \bibfield  {author} {\bibinfo {author} {\bibfnamefont {A.}~\bibnamefont
  {Morgan}}\ and\ \bibinfo {author} {\bibfnamefont {S.}~\bibnamefont {Hogan}},\
  }\bibfield  {title} {\bibinfo {title} {Coupling rydberg atoms to microwave
  fields in a superconducting coplanar waveguide resonator},\ }\href@noop {}
  {\bibfield  {journal} {\bibinfo  {journal} {Physical Review Letters}\
  }\textbf {\bibinfo {volume} {124}},\ \bibinfo {pages} {193604} (\bibinfo
  {year} {2020})}\BibitemShut {NoStop}%
\bibitem [{\citenamefont {Browaeys}\ and\ \citenamefont
  {Lahaye}(2016)}]{browaeys2016interacting}%
  \BibitemOpen
  \bibfield  {author} {\bibinfo {author} {\bibfnamefont {A.}~\bibnamefont
  {Browaeys}}\ and\ \bibinfo {author} {\bibfnamefont {T.}~\bibnamefont
  {Lahaye}},\ }\bibfield  {title} {\bibinfo {title} {Interacting cold rydberg
  atoms: A toy many-body system},\ }in\ \href@noop {} {\emph {\bibinfo
  {booktitle} {Niels Bohr, 1913-2013: Poincar{\'e} Seminar 2013}}}\ (\bibinfo
  {organization} {Springer},\ \bibinfo {year} {2016})\ pp.\ \bibinfo {pages}
  {177--198}\BibitemShut {NoStop}%
\bibitem [{\citenamefont {Arias}\ \emph {et~al.}(2019)\citenamefont {Arias},
  \citenamefont {Lochead}, \citenamefont {Wintermantel}, \citenamefont
  {Helmrich},\ and\ \citenamefont {Whitlock}}]{arias2019realization}%
  \BibitemOpen
  \bibfield  {author} {\bibinfo {author} {\bibfnamefont {A.}~\bibnamefont
  {Arias}}, \bibinfo {author} {\bibfnamefont {G.}~\bibnamefont {Lochead}},
  \bibinfo {author} {\bibfnamefont {T.~M.}\ \bibnamefont {Wintermantel}},
  \bibinfo {author} {\bibfnamefont {S.}~\bibnamefont {Helmrich}},\ and\
  \bibinfo {author} {\bibfnamefont {S.}~\bibnamefont {Whitlock}},\ }\bibfield
  {title} {\bibinfo {title} {Realization of a rydberg-dressed ramsey
  interferometer and electrometer},\ }\href@noop {} {\bibfield  {journal}
  {\bibinfo  {journal} {Physical Review Letters}\ }\textbf {\bibinfo {volume}
  {122}},\ \bibinfo {pages} {053601} (\bibinfo {year} {2019})}\BibitemShut
  {NoStop}%
\end{thebibliography}%
\end{document}